%% file: article.tex
\documentclass[10pt]{sigplanconf}

\usepackage{url}
\usepackage{amsmath}
\usepackage{color}
\usepackage{longtable}
\usepackage{supertabular}
\usepackage{multicol}
\usepackage{lipsum}
\usepackage{array}
\usepackage{listings}
\usepackage{tipa}

\usepackage{tikz} 
\usetikzlibrary{decorations.text}
\usetikzlibrary{patterns}
\usetikzlibrary{calc,trees,positioning,arrows,chains,shapes.geometric,%
    decorations.pathreplacing,decorations.pathmorphing,shapes,%
    matrix,shapes.symbols}
\usepackage[]{hyperref}
\hypersetup{
  colorlinks = true,
    linkcolor = black,
    citecolor = black,
    filecolor = black,
    urlcolor = black,
}

\newcolumntype{H}{>{\setbox0=\hbox\bgroup}c<{\egroup}@{}} 

\newcommand{\toolname}{Dexpler}

\begin{document}

\conferenceinfo{SOAP'12}{June 14, Beijing, China.}
\copyrightyear{2012}
\copyrightdata{ISBN 978-1-4503-1490-9/12/06}


\title{\toolname{}: Converting Android Dalvik Bytecode to Jimple for Static Analysis with Soot}

\authorinfo{Alexandre Bartel \and Jacques Klein \\ \and Yves Le Traon}
           {University of Luxembourg - SnT, Luxembourg}
           {firstname.lastname@uni.lu}
\authorinfo{Martin Monperrus}
           {University of Lille - INRIA, France}
           {martin.monperrus@univ.lille1.fr}

\maketitle

\begin{abstract}
	\input{abstract}
\end{abstract}




\section{Introduction} 
	\label{sec:intro}
	\input{intro}

\section{Soot} 
	\label{sec:soot}
	\input{soot}

\section{Dalvik Bytecode}
	\label{sec:dalvik-bytecode}
	\input{dalvik-bytecode}

\section{\toolname{}} 
	\label{sec:dexpler}
	\input{dexpler}

\section{Evaluation}
	\label{sec:eval}
	\input{eval}

\section{Current Liminations}
	\label{sec:limitations}
	\input{limitations}

\section{Related Work}
	\label{sec:rwork}
	\input{rwork}

\section{Conclusion}
	\label{sec:conclu}
	\input{conclu}

\acks

This research is supported by the National Research Fund, Luxembourg.


\bibliographystyle{abbrvnat}
\bibliography{bib/bib}

\clearpage 
\begin{multicols}{2}  
\end{multicols}				%
\onecolumn						%

\appendix
\section{Jimple Code} 
	\label{appendix:dalvikJimpleTable}

\input{dalvik}

\end{longtable}
\begin{multicols}{2}
\end{multicols}


\end{document}

%% file: abstract.tex
This paper introduces \toolname{}, a software package which converts Dalvik bytecode to Jimple.
\toolname{} is built on top of Dedexer and Soot.
As Jimple is Soot's main internal representation of code, the Dalvik bytecode can be
manipulated with any Jimple based tool, for instance for performing point-to or flow analysis. 

%% file: intro.tex
Android applications are mainly written in Java. However, they are not distributed as Java bytecode but rather as
Dalvik bytecode. Indeed, the original Java code is first compiled into Java bytecode
which is then transformed into Dalvik bytecode by the \emph{dx} tool\footnote{\emph{dx} is part of the Android SDK available at \url{http://developer.android.com/sdk/index.html}}. 
Dalvik bytecode is register based and optimized to run
on devices where memory and processing power are scarce.

Analyzing Android applications with Java static analysis tools means either that the Java source code or the
Java bytecode of the Android application must be available. Most of the time, Android applications developers do not distribute the source code
of their applications. One must then analyze the bytecode, for instance for malware detection.

Thus, to analyse Android applications, one is forced to use a Dalvik disassembler such as Smali \cite{smali} or Androguard \cite{androguard}.
The problem with disassemblers is that they generaly use their own representation of the bytecode which prevents them to use existing tools. 

Another possibility is to first convert Dalvik bytecode to Java bytecode using Ded \cite{Enck:2011:SAA:2028067.2028088}, Dex2jar \cite{dex2jar} or undx \cite{dalvik_undx} 
and then use Java tailored static analysis tools such as Soot \cite{vall99soot}, BCEL \cite{DahmM:BCEL} or WALA \cite{wala}. 
Tools which generate Java bytecode can leverage existing Java bytecode analyzers. However,
the conversion from Dalvik to Java bytecode could be avoided by directly converting Dalvik bytecode to the internal representation of a tool.

We introduce \toolname{}\footnote{\toolname{} webpage: 
\url{http://www.abartel.net/dexpler/}}, a Soot modification which allows Soot to
directly read Dalvik bytecode and perform
analysis and/or transformation on it's internal Jimple representation.
Using this method eliminates the intermediate Dalvik to Java bytecode conversion step and  
enables to use a faster and simpler tool chain for static analysis. \toolname{} only uses
a disassembler and then does the rest of the work itself or by using Soot.

The contributions of this paper are the following:
\begin{itemize}
	\item we describe a Dalvik to Jimple converter tool 
	\item we provide a comprehensive table which maps Dalvik bytecode instructions to Jimple statements
\end{itemize}

The reminder of this paper is organized as follows. In Section \ref{sec:soot} we explain
what Soot is, and how it has been modified to handle Dalvik bytecode.
Section \ref{sec:dalvik-bytecode} is an overview of the Dalvik bytecode.
In Section \ref{sec:dexpler} we propose a Soot modification called \toolname{} which enables Soot to read
Dalvik bytecode. 
In Section \ref{sec:eval} we evaluate \toolname{} on test cases and on one Android application, present and
discuss the results. Section \ref{sec:limitations} explains the current limitation of our tool.
We present the related work in Section \ref{sec:rwork}.
Finally we conclude the paper and discuss open research challenges in Section \ref{sec:conclu}.

%% file: soot.tex
In this Section we give a brief overview of Soot and then describe
how we incorporate \toolname{} in Soot.

\subsection{Soot Overview}

Soot \cite{vall99soot, cetus11soot} was created as a Java compiler testbed at McGill University. It has evolved to become
a Java static analysis and transformation tool. 

Soot can be used as a code analyzer to, among others,  
check that certain properties hold \cite{Yahav:2004:VSP} or guarantee correctness of programs \cite{journals/entcs/Fredlund05}. 

Multiple tools based on Soot have been developed to perform transformations such as translation of Java to C \cite{conf/date/VarmaB04}, instrumentation of Java
programs \cite{conf/icse/ZhangYZCY10}, obfuscator for Java \cite{DRM'03*142}, software watermarking \cite{Cousot:2004:AIB}, ...\footnote{see \url{https://svn.sable.mcgill.ca/wiki/index.cgi/SootUsers} for a comprehensive list}.

Soot accepts Java source code, Java bytecode and Jimple source code as input files.
Whatever the input format, it is converted into Soot's internal representation: Jimple. Java sIMPLE,
is a stack-less, three address representation which features only 15 instructions. Any method code
can be viewed as a graph of Jimple statements associated with a list of Jimple local variables.

\subsection{From Java Bytecode to Jimple}

We now describe how Soot handles Java bytecode classes.
In a typical case, Soot is launched by specifying the target directory as a parameter. This directory contains the code of the program
to analyze, called \texttt{Application Code} (only Java bytecode in this example). 
First, the \texttt{main()} method of the \texttt{Main} class is executed and calls \texttt{Scene.load\nolinebreak[0]Necessary\nolinebreak[0]Classes()}. 
This method loads basic Java classes and then loads specific \texttt{Application classes} by calling \texttt{load\nolinebreak[0]Class()}. Then,
\texttt{Soot\nolinebreak[0]Resolver.resolve\nolinebreak[0]Class()} is called. 
The resolver calls \texttt{Source\nolinebreak[0]Locator.getClass\nolinebreak[0]Source()} to fetch a reference to a \texttt{ClassSource}, an interface between the file containing the Java bytecode and Soot.
In our case the class source is a \texttt{CoffiClassSource} because it is the \emph{coffi} module which handles the conversion from Java bytecode to Jimple.
When the resolver has a reference to a class source, it calls \texttt{resolve()} on it. This methods in turn calls \texttt{soot.coffi.Util.resolve\nolinebreak[0]From\nolinebreak[0]Class\nolinebreak[0]File()} which
creates a \texttt{SootClass} from the corresponding Java bytecode class. All source fields of Soot class' methods are set to refer to a \texttt{CoffiMethodSource} object. 
This object is used later to get the Jimple representation of the method. 

For instance, if during an analysis with Soot the analysis code calls \texttt{SootMethod.getActiveBody()} and the Jimple code of the method was not already generated, 
\texttt{getActiveBody()} will call \texttt{CofficMethodSource.get\nolinebreak[0]Body()} to compute Jimple code from the Java bytecode. The Jimple code representation of the
method can then be analyzed and/or transformed.

\subsection{Soot and Dalvik}

Soot is missing a Dalvik to Jimple transformation module. We implemented such a module called \toolname{} 
and incorporated it to Soot using the same structure as Soot's Java bytecode parser module, \emph{coffi}
by adding the \texttt{DalvikClassSource} and {DalvikMethodSource} classes.

%% file: dalvik-bytecode.tex
We present in this Section the structure of a \texttt{.dex} file containing Dalvik classes and Dalvik bytecode.

\subsection{Overall Structure}

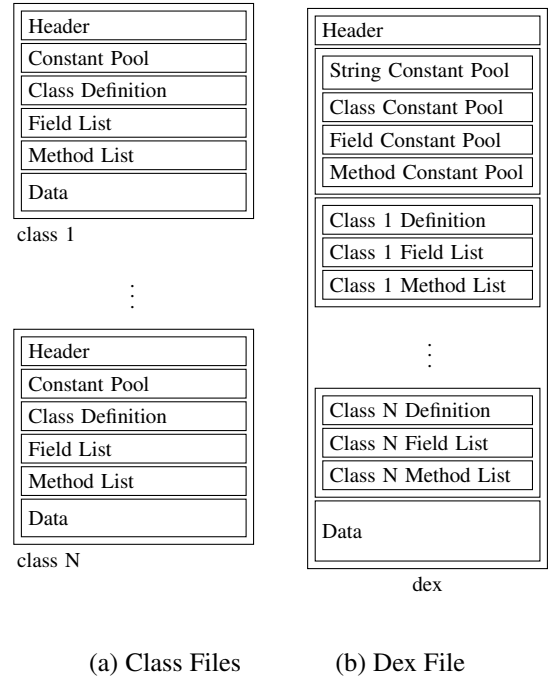
\begin{figure}[!]
\begin{center}
\footnotesize
  \input{figures/dalvikVsJava}                                                                                                                                               
\caption{\label{fig:dalvikVsJava}Dalvik Dex and Java Class}
\end{center}
\end{figure}

A single Dalvik executable is produced from $N$ Java bytecode classes through the \emph{dx} compiler.
The resulting Dalvik bytecode is stored in a \texttt{.dex} file as represented in Figure \ref{fig:dalvikVsJava}b. 

As represented in Figure \ref{fig:dalvikVsJava}a, there is only a single place where literal constant values are stored (constant pool) per Java class. 
It is heterogeneous since different kind of Objects are mixed together (ex: Class, MethodRef, Integer, String, ...). 
A \texttt{.dex} file contains four homogeneous constants pools: for Strings, Class, Fields and Methods.
It is shared by all the classes. A \texttt{.dex} file contains multiple \emph{Class Definitions} each containing one or more \emph{Method definition}
each of those being linked to Dalvik bytecode instructions present in the \emph{Data} section.

\subsection{Dalvik Instruction}

The Dalvik virtual machine is register based. This means most instructions must specify the registers which they manipulate. Registers
could be specified on 4, 8 or 16 bits depending on the instruction.

There are 237 opcodes present in the Dalvik opcode constant list\footnote{dalvik/bytecode/Opcodes.java}.
However, 12 odex (optimized dex) instructions can not be found in Android applications Dalvik bytecode as they are unsafe instructions 
generated within the Android system to optimize Dalvik bytecode. 
Moreover, 8 instructions were never found in application code \cite{url:dalvikOpcodes}. 
According to those numbers, only 217 instructions can be found in Android PacKages (.apk) in practice.

The set of instructions can be divided between instructions which provide the type of the registers they manipulate (ex: \texttt{sub-long v1, v2, v3}) 
and those which do not (ex: \texttt{const v0, 0xBEEF}). Moreover, there is no distinction between \texttt{NULL} and \texttt{0} which are both represented
as \texttt{0} (see Figure \ref{fig:zeronullbytecode-ex}). As we will see in Section \ref{sec:dexpler}, the lack of type and the \texttt{NULL} representation  become problematic 
when translating the Dalvik bytecode to Jimple.

\begin{figure}
\footnotesize
\begin{verbatim}
int i = 0;                      00: const/4 v0, #int 0
Object o = null;                01: const/4 v1, #int 0 
(Java)                          (Dalvik)
\end{verbatim}
\caption{Dalvik Representation of \emph{null} and \emph{zero}}
\label{fig:zeronullbytecode-ex}
\end{figure}

%% file: figures/dalvikVsJava.tex
\newcommand{\UCFPSOFFSET}{5.4}
\newcommand{\POLICYOFFSET}{5.6}

\pgfdeclarelayer{background}
\pgfdeclarelayer{foreground}
\pgfsetlayers{background,main,foreground}
\tikzstyle{block} = [draw,fill=white!20,minimum width=4em]
\tikzstyle{steps} = [circle, draw, above, inner sep=1pt, midway, yshift=.2cm, fill=black!50, text=white]
\begin{tikzpicture} [node distance=.5cm, start chain=going below, remember picture]

		  \tikzstyle{vertex}=[circle,fill=black!25,minimum size=14pt,inner sep=0pt]
		  \tikzstyle{vInterface}=[block,fill=black!25,minimum size=14pt,inner sep=0pt]
		  \tikzstyle{vPermission}=[diamond,fill=blue!25,minimum size=14pt,inner sep=0pt]
		  \tikzstyle{bytecode}=[]
  		\tikzstyle{tuborg}=[decorate]
  		\tikzstyle{tubnode}=[midway, right=2pt]
			\tikzstyle{stub} = [rectangle, fill=yellow!5, text width=1.8cm, minimum height=.5cm, node distance=.6cm, draw];
			\tikzstyle{filebox} = [rectangle, text width=3cm, draw, minimum height=2.3cm, node distance=4cm];
			\tikzstyle{insideBox} = [rectangle, fill=white, text width=2.8cm, minimum height=.3cm, draw, node distance=.2cm];
			\tikzstyle{insideInsideBox} = [rectangle, fill=white, text width=2.6cm, minimum height=.3cm, draw, node distance=.2cm];

		\node[text width=3.1cm] (Box-classfiles) at (-2.5,0){
			\tikz \node[filebox, label=below: class 1] (class1) {
				\tikz \node[insideBox] {Header};
				\tikz \node[insideBox] {Constant Pool};
				\tikz \node[insideBox] {Class Definition};
				\tikz \node[insideBox] {Field List};
				\tikz \node[insideBox] {Method List};
				\tikz \node[insideBox, minimum height=.5cm] {Data};
			};
			\tikz \node[minimum height=1cm] (class2) {
				\hspace{1.5cm}\vdots
			};
			\tikz \node[filebox, label=below: class N] (class3) {
				\tikz \node[insideBox] {Header};
				\tikz \node[insideBox] {Constant Pool};
				\tikz \node[insideBox] {Class Definition};
				\tikz \node[insideBox] {Field List};
				\tikz \node[insideBox] {Method List};
				\tikz \node[insideBox, minimum height=.5cm] {Data};
			};
		};

		\node[ filebox, right of=Box-classfiles, label=below:dex] (Box-dexfile) {
			\tikz \node[insideBox] {Header};
			\tikz \node[insideBox] {
				\tikz \node[insideInsideBox] {String Constant Pool};
				\tikz \node[insideInsideBox] {Class Constant Pool};
				\tikz \node[insideInsideBox] {Field Constant Pool};
				\tikz \node[insideInsideBox] {Method Constant Pool};
			};
			\tikz \node[insideBox] {
				\tikz \node[insideInsideBox] {Class 1 Definition};
				\tikz \node[insideInsideBox] {Class 1 Field List};
				\tikz \node[insideInsideBox] {Class 1 Method List};
			};
			\tikz \node[minimum height=1cm] {\hspace{1.4cm}\vdots};
			\tikz \node[insideBox] {
				\tikz \node[insideInsideBox] {Class N Definition};
				\tikz \node[insideInsideBox] {Class N Field List};
				\tikz \node[insideInsideBox] {Class N Method List};
			};
			\tikz \node[insideBox, minimum height=.8cm] {Data};
		};
	\node[text width = 5cm] at (-.5,-5){\normalsize (a) Class Files \hfill (b) Dex File};

\begin{pgfonlayer}{background}
\end{pgfonlayer}
\end{tikzpicture}

%% file: dexpler.tex
This section describes Dexpler, the Dalvik to Jimple converter tool. It leverages the \emph{dedexer} \cite{dedexer} 
Dalvik bytecode disassembler and the Soot \emph{fast typing} Jimple component 
implementing a type inferrence algorithm \cite{conf/oopsla/BellamyAMS08} for local variables.
We first give a brief overview on \emph{dedexer} and on how \toolname{} is working in Sections \ref{sec:dedexer} and \ref{sec:derOverview}, respectively. 
Then, we detail issues we have to deal with. 

\subsection{Dedexer}
	\label{sec:dedexer}

Our tool leverages \emph{dedexer} a Dalvik bytecode parser and disassembler which generates Jasmin \cite{jasmin, url:jasmin} like text files containing
Dalvik instructions instead of Java instructions.
We generate Jimple classes, methods and statements from the informations provided by \emph{dedexer}'s dex file parser.

\subsection{Overview}
	\label{sec:derOverview}
Dalvik bytecode instructions are first mapped to Jimple statements and
registers mapped to Jimple local variables. The type of local variables is set to \texttt{UnknownType}. 
Then, Soot's Jimple component, \emph{fast typing}, is applied to infer the type of the local variables.
The third and last step consists in applying Soot's Jimple pack \emph{jop}, which features components such as \emph{nop eleminitor}, to optimize the generated Jimple code.

\subsection{Instruction Mapping}

Each Dalvik instruction is mapped to a corresponding (or a group of) Jimple statements. A comprehensive mapping is
represented in Table \ref{table:dalvikJimple} in Appendix \ref{appendix:dalvikJimpleTable}. Unused opcodes are
marked as '\emph{-}' and odex opcodes as '\emph{odex}'. There are five main groups of instructions: move instructions (0x01 to 0x1C), 
branch instructions (0x27 to 0x3D), getter and setter instructions (0x44 to 0x6D), method invoke instructions (0x6E to 0x78) logic
and arithmetic instructions (0x7B to 0xE2).

\subsection{Type Inference}
	\label{sec:typeInference}

The type for local variables is inferred using the \emph{fast typing} Soot component.
However, the inference algorithm sometime generates an exception and stop
because some Dalvik instructions (such as the constant initialization instructions 
\texttt{0x12} to \texttt{0x19}) do not provide enough information and thus confuse
the inference engine.

The lack of type is present in the following instructions:
\begin{itemize}
	\item null initialization instructions (zero or null?)
	\item initialization instructions (32 bits: integer or float?, 64 bits: long or double?)
\end{itemize}

\paragraph{Null Initialization} 
Figure \ref{fig:zeronullbytecode} illustrates the problem with a bytecode snippet generated from
the Java code of Figure \ref{fig:zeronullcode}. Register \texttt{v0} is initialized with 0 at \texttt{01}.
At this point we do not know if \texttt{v0} is an integer, a float or a reference to an object.
At \texttt{02} we still do not have the answer. We have to wait until instruction at \texttt{04}
to known that the type of \texttt{v0} is \texttt{Coordinate}. At this point, the Jimple instruction
generated for \texttt{01} has to be updated to use \texttt{NullConstant} instead
of the default \texttt{IntConstant(0)}. If this is not handled correctly, the \emph{fast typing} component 
generates an exception and stops.

\paragraph{Numeric Constant Initialization}
Similarly, \emph{float} constants initialization cannot be distinguished from \emph{int} constants initialization
and \emph{double} constants initialization from \emph{long} constants initialization. Thus, we go through the
graph of Jimple statements to find how constants are used and correct the initializations Jimple statements when needed.
For instance, if a \emph{float/int} constant (initialized by default to \emph{int} in the Jimple statement) is
later used in a \emph{float} addition, the constant initialization changes from \texttt{IntConstant(c)} to
\texttt{Float\nolinebreak[0]Constant(\nolinebreak[0]Float.\nolinebreak[0]int\nolinebreak[0]Bits\nolinebreak[0]To\nolinebreak[0]Float(\nolinebreak[0]c))}.

\paragraph{}
We implemented the algorithm described by Enck et al. \cite{Android-ded}. It is based on algorithms which extract typing information for a variable by
looking at how it is used in operations whithin which the type of the operands is knows (ex: the variable is used as a parameter of a method
invocation) \cite{Milner78, DBLP:conf/mfcs/Tiuryn90}.
For each ambiguous register declaration, the algorithm performs a depth first search in the control flow graph of Jimple statements
to find out how the declared local variable \emph{dv} (registers are mapped to Jimple local variables) is used. The type of \emph{dv}
is exposed with the following statements: comparison with a known type, instructions operating only on specific types (ex: neg-float),
non-void return instructions and method invocation.
The search in a branch of the graph is terminated if either the local variable is reassigned (new declaration) or if there is no more statement that
follow the current one (eg: the current statement is a return or throw statement).
When the type information is found it is forward propagated to all subsequent ambiguous uses between the target ambiguous declaration of \emph{dv}
and any new declaration of \emph{dv}.

\begin{figure}
\footnotesize
\begin{verbatim}
Coordinate newCoord = null;
while (newCoord!=null) {
 newCoord = new Coordinate(1,1);
}   
if (newCoord == null) {
 [...]
}   
\end{verbatim}
\caption{Illustration of the \emph{null} init problem.}
\label{fig:zeronullcode}
\end{figure}

\begin{figure}
\footnotesize
\begin{verbatim}
00: const/4 v1, #int 1 
01: const/4 v0, #int 0
02: if-eqz v0, 000a
04: new-instance v0, LCoordinate;
06: invoke {v0, v1, v1}, LCoordinate;.<init>:(II)V
09: goto 0002
0a: if-nez v0, 0013
[...]
13: ...
\end{verbatim}
\caption{Resulting Dalvik Bytecode from Figure \ref{fig:zeronullcode}}
\label{fig:zeronullbytecode}
\end{figure}

\subsection{Handling Branches}
	\label{sec:branches}

Dalvik instructions are mapped to Jimple statements.
When parsing Dalvik bytecode, we keep a mapping between bytecode instructions addresses and Jimple statements. 
Thus, when a Dalvik branch instruction is parsed, a Jimple jump instruction is generated and  its target is retrieved by fetching
the Jimple statement mapped to the Dalvik branch instruction target's address. 
We add a \texttt{nop} instruction as the first instruction of every Jimple methods.
This way, if the first Dalvik instruction is a jump or if the jump's target correspond to a non-yet generated Jimple statement, we redirect it to the 
this \texttt{nop} Jimple instruction. We correct those Jimple jump instructions once the whole Dalvik bytecode of the method has been processed: at this
point we know the target Jimple statement mapped to the Dalvik jump's target address.
The Jimple \texttt{nop} instruction we add is removed during the Jimple optimization step.

Branching instructions often rely on the result of a comparison of two registers. Dalvik comparisons
between \emph{double} or \emph{float} are explicit and provide typing information.
However, when a register $r$ is compared with zero one has to check the type of $r$. If it
is an object, we change the zero value to \emph{null} since it is a comparison between objects.
We do this change when the \emph{fast typing} component has finished.
Indeed, comparisons do not influence the type inference.
For example, the Jimple statement generated from \texttt{02} in Figure \ref{fig:zeronullbytecode} 
has to be updated to use \texttt{NullConstant} instead of \texttt{IntConstant(0)}.
If this is not handled correctly the bytecode generated from Jimple statements
does not run correctly and generates an exception similar to the following one: \texttt{Exception in thread "main" java.lang.VerifyError:} \texttt{Expecting to find integer on stack}.

\paragraph{}
Dexpler enables us to transform Dalvik bytecode to Jimple representation. From there, Soot can 
be used as a static analysis tool to analyze the code.
The next Section evaluates \toolname{}.

%% file: eval.tex
We evaluate \toolname{} using test cases, and one Android application: Snake.

\subsection{Test Cases}

The first step is to generate the Dalvik bytecode for every test case.
The test cases are written in Java then compiled into Java bytecode using \emph{javac} and finally converted
into Dalvik bytecode using \emph{dx}. 
The second step is to execute \toolname{} on every generated Dalvik bytecode test case.
This generates \texttt{.jimple} and \texttt{.class} files. We then compare the execution result
from of the versions produced from the original Java bytecode and the Java bytecode produced by Soot
from the Dalvik bytecode. 
Executions of the \texttt{.class} files give the correct result.

We wrote test cases for arithmetic operations, branches, method calls, array initialization, string manipulation, null 
and zero usage, exceptions and casts.

Since simple test cases do not reflect a real application we also evaluated our tool on one Android application.

\subsection{Android Application}

The snake application is a demonstration application developed by the Android team to showcase the Android platform.\footnote{\url{http://developer.android.com/resources/samples/Snake/index.html}}
It features 11 classes, 39 methods and was written in 550 lines of Java code. 
The generated Dalvik bytecode takes 14 KiB and contains 884 Dalvik instructions.

From the Dalvik bytecode of the Snake application we generate Jimple code in one second (duration for the Dalvik to Jimple conversion only). 
Then we ask Soot to generate Java bytecode from the Jimple representation. We convert the Java bytecode
back to Dalvik, repackage an Android application and launch it on the Android emulator.

The application runs smoothly and the game is working. 

\subsection{Static Analysis on Snake}

\begin{figure}
	\begin{center}
		\includegraphics[width=8cm]{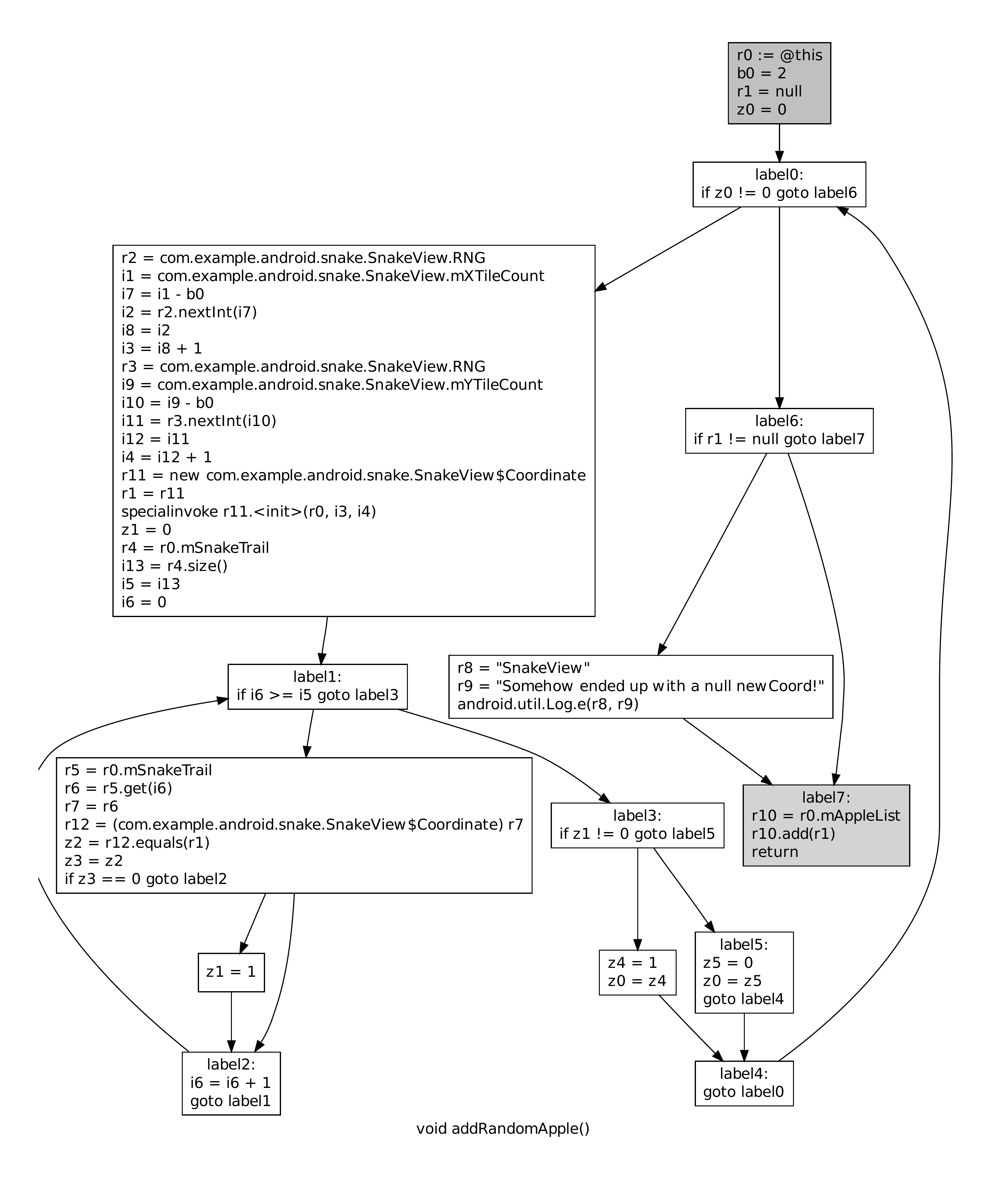}
	\end{center}
	\caption{Control Flow Graph for \texttt{addRandomApple} Method Extracted from the Generated Jimple Representation.}
	\label{fig:cfgSnake}
\end{figure}

We use Soot to generate a call graph of the Snake application as well as a control flow graph represented in Figure \ref{fig:cfgSnake} in 14 seconds 
(duration from the launch time of Soot until Soot has finished).
We perform this to check that the generated call graph and CFG correspond to the original code meaning that the conversion from Dalvik to Jimple
is correct for this code.

\paragraph{}
We have successfully tested our prototype tool on test cases as well as on an Android application.

%% file: limitations.tex
The current version of \toolname{} is able to transform Android applications such as the Snake game. 

However, it does not handle optimized Dalvik (odex) opcodes.

Moreover, when inferring types for ambiguous declarations the algorithm supposes
that the Dalvik bytecode is legal in the sense that it was generated from Java
source code and not hand-crafted by malicious developers. In the later case 
assumptions such as "comparisons always involve variables with the same type" may
not hold anymore and may make \toolname{} to infer wrong types.

%% file: rwork.tex
To our knowledge no existing tool directly converts Dalvik bytecode to Jimple.
We either found tools to convert Dalvik bytecode to Java bytecode or tools
to disassemble and/or assemble Dalvik bytecode using an intermediate representation.

\paragraph{Dalvik to Java Bytecode Converter}
Ded \cite{Enck:2011:SAA:2028067.2028088} is a Dalvik bytecode to Java bytecode converter. Once the Java
bytecode is generated, Soot is used to optimize the code. 
Dex2jar \cite{dex2jar} also generates Java bytecode from Dalvik bytecode but no not use any external
tool to optimize the resulting Java bytecode.
Undx \cite{dalvik_undx} is also a Dalvik to Java bytecode converter but seems to be unavailable.

We on the other hand do not directly generate Java bytecode but Jimple code. 
From there, since the Jimple code is within Soot, we can generate Java bytecode as well.

\paragraph{Dalvik Assembler/Disassembler}
Smali \cite{smali} or Androguard \cite{androguard} can be used to reverse engineer Dalvik bytecode.
They use their own representation of the Dalvik bytecode: they can not leverage existing analysis tools.

Our tool, use Soot's internal representation which allows existing tools to analyze/transform the Dalvik bytecode.

%% file: conclu.tex
We have presented \toolname{}\footnote{\toolname{} webpage: 
\url{http://www.abartel.net/dexpler/}} a Soot modification with enables Soot to
analyse Dalvik bytecode and thus Android applications. This tool leverages
\emph{dedexer} for the parsing of Dalvik \texttt{dex} 
files and Soot's \emph{fast typing} component for the type inference.

\toolname{} converts every Dalvik instruction to Jimple. We are working on improving \toolname{} 
to make it robust to yet unhandled typing issues. Once this step is done we will look at the performance
of this tool compared to current Java bytecode generation and analysis tools.

%% file: dalvik.tex
\begin{longtable}{>{\centering\arraybackslash}m{1.5cm}|p{4.5cm}HHH|p{10cm}}
\caption{Jimple Code representation of Dalvik Instructions} \\
\label{table:dalvikJimple}
Opcode & Opcode name & Explanation & Example & Jimple Statements & Jimple Code \\ \hline \hline
\endhead
0x00 & nop & No operation & 0000 - nop & newNopStmt () & nop \\ \hline
0x01 & move vx,vy & Moves the content of vy into vx. Both registers must be in the first 256 register range. & 0110 - move v0, v1  Moves v1 into v0. & newAssignStmt (vx, vy) & vx = vy \\ \hline
0x02 & move/from16 vx,vy & Moves the content of vy into vx. vy may be in the 64k register range while vx is one of the first 256 registers. & 0200 1900 - move/from16 v0, v25  Moves v25 into v0. & newAssignStmt (vx, vy) & vx = vy \\ \hline
0x03 & move/16 &  & & newAssignStmt (vx, vy) & vx = vy \\ \hline
0x04 & move-wide  &  & & newAssignStmt (vx, vy) & vx = vy \\ \hline
0x05 & move-wide/from16 vx,vy & Moves a long/double value from vy to vx. vy may be in the 64k register range while wx is one of the first 256 registers. & 0516 0000 - move-wide/from16 v22, v0  Moves v0 into v22. & newAssignStmt (vx, vy) & vx = vy \\ \hline
0x06 & move-wide/16 &  & & newAssignStmt (vx, vy) & vx = vy \\ \hline
0x07 & move-object vx,vy & Moves the object reference from vy to vx. & 0781 - move-object v1, v8 Moves the object reference in v8 to v1. & newAssignStmt (vx, vy) & vx = vy \\ \hline
0x08 & move-object/from16 vx,vy & Moves the object reference from vy to vx, vy can address 64k registers and vx can address 256 registers. & 0801 1500 - move-object/from16 v1, v21  Move the object reference in v21 to v1. & newAssignStmt (vx, vy) & vx = vy \\ \hline
0x09 & move-object/16 &  & & newAssignStmt (vx, vy) & vx = vy \\ \hline
0x0A & move-result vx & Move the result value of the previous method invocation into vx. & 0A00 - move-result v0  Move the return value of a previous method invocation into v0. & newAssignStmt (vx, mres) & vx = mres \\ \hline
0x0B & move-result-wide vx & Move the long/double result value of the previous method invocation into vx,vx+1. & 0B02 - move-result-wide v2  Move the long/double result value of the previous method invocation into v2,v3. & newAssignStmt (vx, mres) & vx = mres \\ \hline
0x0C & move-result-object vx & Move the result object reference of the previous method invocation into vx. & 0C00 - move-result-object v0 & newAssignStmt (vx, mres) & vx = mres \\ \hline
0x0D & move-exception vx & Move the exception object reference thrown during a method invocation into vx.  & 0D19 - move-exception v25 & & vx = mres \\ \hline
0x0E & return-void & Return without a return value & 0E00 - return-void & newReturnVoidStmt () & return \\ \hline
0x0F & return vx & Return with vx return value & 0F00 - return v0  Returns with return value in v0. & newReturnStmt (vx) & return vx \\ \hline
0x10 & return-wide vx & Return with double/long result in vx,vx+1. & 1000 - return-wide v0  Returns with a double/long value in v0,v1. & newReturnStmt (vx) & return vx \\ \hline
0x11 & return-object vx & Return with vx object reference value. & 1100 - return-object v0  Returns with object reference value in v0 & newReturnStmt (vx) & return vx \\ \hline
0x12 & const/4 vx,lit4 & Puts the 4 bit constant into vx & 1221 - const/4 v1, \#int2  Moves literal 2 into v1. The destination register is in the lower 4 bit in the second byte, the literal 2 is in the higher 4 bit. & newAssignStmt (vx, IntConstant.v (lit4))& vx = lit4 \\ \hline
0x13 & const/16 vx,lit16 & Puts the 16 bit constant into vx & 1300 0A00 - const/16 v0, \#int 10  Puts the literal constant of 10 into v0. & newAssignStmt (vx, IntConstant.v (lit16)) & vx = lit16 \\ \hline
0x14 & const vx, lit32 & Puts the integer constant into vx & 1400 4E61 BC00 - const v0, \#12345678 // \#00BC614E Moves literal 12345678 into v0. & & vx = lit32 \\ \hline
0x15 & const/high16 v0, lit16 & Puts the 16 bit constant into the topmost bits of the register. Used to initialize float values. & 1500 2041 - const/high16 v0, \#float 10.0 // \#41200000 Moves the floating literal of 10.0 into v0. The 16 bit literal in the instruction carries the top 16 bits of the floating point number. & newAssignStmt (vx, FloatConstant.v (Float.intBitsToFloat (lit16 $<<$ 16)))& vx = lit16 $<<$ 16 \\ \hline
0x16 & const-wide/16 vx, lit16 & Puts the integer constant into vx and vx+1 registers, expanding the integer constant into a long constant.. & 1600 0A00 - const-wide/16 v0, \#long 10 Moves literal 10 into v0 and v1 registers. & newAssignStmt (vx, LongConstant.v (lit16)) & vx = lit16 \\ \hline
0x17 & const-wide/32 vx, lit32 & Puts the 32 bit constant into vx and vx+1 registers, expanding the integer constant into a long constant. & 1702 4e61 bc00 - const-wide/32 v2, \#long 12345678 // \#00bc614e Puts \#12345678 into v2 and v3 registers. & newAssignStmt (vx, LongConstant.v (lit32)) & vx = lit32 \\ \hline
0x18 & const-wide vx, lit64 & Puts the 64 bit constant into vx and vx+1 registers. & 1802 874b 6b5d 54dc 2b00- const-wide v2, \#long 12345678901234567 // \#002bdc545d6b4b87 Puts \#12345678901234567 into v2 and v3 registers. & newAssignStmt (vx, DoubleConstant.v (lit64)) & vx = lit64 \\ \hline
0x19 & const-wide/high16 vx,lit16 & Puts the 16 bit constant into the highest 16 bit of vx and vx+1 registers. Used to initialize double values. & 1900 2440 - const-wide/high16 v0, \#double 10.0 // \#402400000 Puts the double constant of 10.0 into v0 register. & newAssignStmt (vx, DoubleConstant.v (Double.longBitsToDouble(lit16$<<$48))) & vx = lit16 $<<$ 48 \\ \hline
0x1A & const-string vx,string\_id & Puts reference to a string constant identified by string\_id into vx. & 1A08 0000 - const-string v8, """" // string@0000 Puts reference to string@0000 (entry \#0 in the string table) into v8. & newAssignStmt (vx, string\_id) & vx = string \\ \hline
0x1B & const-string-jumbo vx,string&  & & & vx = string  \\ \hline
0x1C & const-class vx,type\_id & Moves the class object of a class identified by type\_id (e.g. Object.class) into vx. & 1C00 0100 - const-class v0, Test3 // type@0001 Moves reference to Test3.class (entry\#1 in the type id table) into  & & vx = class "type" \\ \hline
0x1D & monitor-enter vx & Obtains the monitor of the object referenced by vx. & 1D03 - monitor-enter v3 Obtains the monitor of the object referenced by v3. & newEnterMonitorStmt (vx) & monitorenter vx \\ \hline
0x1E & monitor-exit vx & Releases the monitor of the object referenced by vx. & 1E03 - monitor-exit v3 Releases the monitor of the object referenced by v3. & newExitMonitorStmt (vx)& monitorexit vx \\ \hline
0x1F & check-cast vx, type\_id & Checks whether the object reference in vx can be cast to an instance of a class referenced by type\_id. Throws ClassCastException if the cast is not possible, continues execution otherwise. & 1F04 0100 - check-cast v4, Test3 // type@0001 Checks whether the object reference in v4 can be cast to type@0001 (entry \#1 in the type id table) & & checkcast = (type) vx \\ \hline
0x20 & instance-of vx,vy,type\_id & Checks whether vy is instance of a class identified by type\_id. Sets vx non-zero if it is, 0 otherwise. & 2040 0100 - instance-of v0, v4, Test3 // type@0001 Checks whether the object reference in v4 is an instance of type@0001 (entry \#1 in the type id table). Sets v0 to non-zero if v4 is instance of Test3, 0 otherwise. & &  vx = vy instanceof type \\ \hline
0x21 & array-length vx,vy & Calculates the number of elements of the array referenced by vy and puts the length value into vx. & 2111 - array-length v1, v1 Calculates the number of elements of the array referenced by v1 and puts the result into v1. & newAssignStmt (vx, newLengthExpr (vy)) & vx = length(vy) \\ \hline
0x22 & new-instance vx,type & Instantiates an object type and puts the reference of the newly created instance into vx. & 2200 1500 - new-instance v0, java.io.FileInputStream // type@0015 Instantiates type@0015 (entry \#15H in the type table) and puts its reference into v0. & newAssignStmt (vx, newNewExpr (soot.RefType.v (type)) & vx = new type \\ \hline
0x23 & new-array vx,vy,type\_id & Generates a new array of type\_id type and vy element size and puts the reference to the array into vx. & 2312 2500 - new-array v2, v1, char[] // type@0025 Generates a new array of type@0025 type and v1 size and puts the reference to the new array into v2. & newAssignStmt (vx, newNewArrayExpr (type\_id, vy)) & vx = new type[vy] \\ \hline
0x24 & filled-new-array \{parameters\},type\_id & Generates a new array of type\_id and fills it with the parameters5. Reference to the newly generated array can be obtained by a move-result-object instruction, immediately following the filled-new-array instruction. & 2420 530D 0000 - filled-new-array \{v0,v0\},[I // type@0D53 Generates a new array of type@0D53. The array's size will be 2 and both elements will be filled with the contents of v0 register. & & vx = new array\_type[size]; vx[0] = e1; ... vx[N] = eN; \\ \hline
0x25 & filled-new-array-range \{vx..vy\},type\_id & Generates a new array of type\_id and fills it with a range of parameters. Reference to the newly generated array can be obtained by a move-result-object instruction, immediately following the filled-new-array instruction.  & 2503 0600 1300 - filled-new-array/range \{v19..v21\}, [B // type@0006 Generates a new array of type@0D53. The array's size will be 3 and the elements will be filled using the v19,v20 and v21 registers4. & & vx = new array\_type[size]; vx[0] = e1; ... vx[N] = eN; \\ \hline
0x26 & fill-array-data vx,array\_data\_offset & Fills the array referenced by vx with the static data. The location of the static data is the sum of  the position of the current instruction and the offset & 2606 2500 0000 - fill-array-data v6, 00e6 // +0025 Fills the array referenced by v0 with the static data at current instruction+25H words location. The offset is expressed as a 32-bit number. The static data is stored in the following format: 0003 // Table type: static array data 0400 // Byte per array element (in this case, 4 byte integers) 0300 0000 // Number of elements in the table 0100 0000  // Element \#0: integer 1 0200 0000 // Element \#1: integer 2 0300 0000 // Element \#2: integer3 & & vx[0] = e1; ... vx[N] = eN; \\ \hline
0x27 & throw vx & Throws an exception object. The reference of the exception object is in vx. & 2700 - throw v0 Throws an exception. The exception object reference is in v0. & newThrowStmt (vx) & throw vx \\ \hline
0x28 & goto target & Unconditional jump by short offset2.  & 28F0 - goto 0005 // -0010 Jumps to current position-16 words (hex 10). 0005 is the label of the target instruction. & newGotoStmt (targetUnit) & goto target \\ \hline
0x29 & goto/16 target & Unconditional jump by 16 bit offset2. & 2900 0FFE - goto/16 002f // -01f1 Jumps to the current position-1F1H words. 002F is the label of the target instruction. & newGotoStmt (targetUnit) & goto target \\ \hline
0x2A & goto/32 target &  & & newGotoStmt (targetUnit) & goto target \\ \hline
0x2B & packed-switch vx,table & Implements a switch statement where the case constants are close to each other. The instruction uses an index table. vx indexes into this table to find the offset of the instruction for a particular case. If vx falls out of the index table, the execution continues on the next instruction (default case). & 2B02 0C00 0000 - packed-switch v2, 000c // +000c Execute a packed switch according to the switch argument in v2. The position of the index table is at current instruction+0CH words. The table looks like the following: 0001 // Table type: packed switch table 0300 // number of elements 0000 0000 // element base 0500 0000  0: 00000005 // case 0: +00000005 0700 0000  1: 00000007 // case 1: +00000007 0900 0000  2: 00000009 // case 2: +00000009 & & switch (vx) \{ case C1: goto target1; ... case CN: goto targetN; \} \\ \hline
0x2C & sparse-switch vx,table & Implements a switch statement with sparse case table. The instruction uses a lookup table with case constants and offsets for each case constant. If there is no match in the table, execution continues on the next instruction (default case). & 2C02 0c00 0000 - sparse-switch v2, 000c // +000c Execute a sparse switch according to the switch argument in v2. The position of the lookup table is at current instruction+0CH words. The table looks like the following. 0002 // Table type: sparse switch table 0300 // number of elements 9cff ffff // first case: -100 fa00 0000 // second case constant: 250 e803 0000 // third case constant: 1000 0500 0000 // offset for the first case constant: +5 0700 0000 // offset for the second case constant: +7 0900 0000 // offset for the third case constant: +9 & & switch (vx) \{ case C1: goto target1; ... case CN: goto targetN; \} \\ \hline
0x2D & cmpl-float & Compares the float values in vy and vz and sets the integer value in vx accordingly3  & 2D00 0607 - cmpl-float v0, v6, v7 Compares the float values in v6 and v7 then sets v0 accordingly. NaN bias is less-than, the instruction will return -1 if any of the parameters is NaN.  & newAssignStmt (vx, newCmplExpr (vy, vz))& vx = vy cmpl vz \\ \hline
0x2E & cmpg-float vx, vy, vz & Compares the float values in vy and vz and sets the integer value in vx accordingly3. & 2E00 0607 - cmpg-float v0, v6, v7 Compares the float values in v6 and v7 then sets v0 accordingly. NaN bias is greater-than, the instruction will return 1 if any of the parameters is NaN.  & newAssignStmt (vx, newCmpgExpr (vy, vz)) & vx = vy cmpg vz \\ \hline
0x2F & cmpl-double vx,vy,vz & Compares the double values in vy and vz2 and sets the integer value in vx accordingly3.  & 2F19 0608 - cmpl-double v25, v6, v8 Compares the double values in v6,v7 and v8,v9 and sets v25 accordingly. NaN bias is less-than, the instruction will return -1 if any of the parameters is NaN.  & newAssignStmt (vx, newCmplExpr (vy, vz)) & vx = vy cmpl vz \\ \hline
0x30 & cmpg-double vx, vy, vz & Compares the double values in vy and vz2 and sets the integer value in vx accordingly3. & 3000 080A - cmpg-double v0, v8, v10 Compares the double values in v8,v9 and v10,v11 then sets v0 accordingly. NaN bias is greater-than, the instruction will return 1 if any of the parameters is NaN.  & newAssignStmt (vx, newCmpgExpr (vy, vz)) & vx = vy cmpg vz \\ \hline
0x31 & cmp-long vx, vy, vz & Compares the long values in vy and vz and sets the integer value in vx accordingly3. & 3100 0204 - cmp-long v0, v2, v4 Compares the long values in v2 and v4 then sets v0 accordingly. & newAssignStmt (vx, newCmpExpr (vy, vz)) & vx = vy cmp vz \\ \hline
0x32 & if-eq vx,vy,target & Jumps to target if vx==vy2. vx and vy are integer values. & 32b3 6600 - if-eq v3, v11, 0080 // +0066 Jumps to the current position+66H words if v3==v11. 0080 is the label of the target instruction. & newIfStmt (newEqExpr (vx, vy), targetUnit)& if (vx == vy) goto target; \\ \hline
0x33 & if-ne vx,vy,target & Jumps to target if vx!=vy2. vx and vy are integer values.  & 33A3 1000 - if-ne v3, v10, 002c // +0010 Jumps to the current position+10H words if v3!=v10. 002c is the label of the target instruction. & newIfStmt (newNeExpr (vx, vy), targetUnit) & if (vx != vy) goto target; \\ \hline
0x34 & if-lt vx,vy,target & Jumps to target is vx<vy2. vx and vy are integer values. & 3432 CBFF - if-lt v2, v3, 0023 // -0035 Jumps to the current position-35H words if v2<v3. 0023 is the label of the target instruction. & newIfStmt (newLtExpr (vx, vy), targetUnit) & if (vx $<$ vy) goto target; \\ \hline
0x35 & if-ge vx, vy,target & Jumps to target if vx>=vy2. vx and vy are integer values. & 3510 1B00 - if-ge v0, v1, 002b // +001b Jumps to the current position+1BH words if v0>=v1. 002b is the label of the target instruction. & newIfStmt (newGeExpr (vx, vy), targetUnit) & if (vx $>$= vy) goto target;  \\ \hline
0x36 & if-gt vx,vy,target & Jumps to target if vx>vy2. vx and vy are integer values. & 3610 1B00 - if-ge v0, v1, 002b // +001b Jumps to the current position+1BH words if v0>v1. 002b is the label of the target instruction. & newIfStmt (newGtExpr (vx, vy), targetUnit) & if (vx $>$ vy) goto target; \\ \hline
0x37 & if-le vx,vy,target & Jumps to target if vx<=vy2. vx and vy are integer values.  & 3756 0B00 - if-le v6, v5, 0144 // +000b Jumps to the current position+0BH words if v6<=v5. 0144 is the label of the target instruction. & newIfStmt (newLeExpr (vx, vy), targetUnit)  & if (vx $<$= vy) goto target; \\ \hline
0x38 & if-eqz vx,target & Jumps to target if vx==02. vx is an integer value. & 3802 1900 - if-eqz v2, 0038 // +0019 Jumps to the current position+19H words if v2==0. 0038 is the label of the target instruction. & newIfStmt (newEqExpr (vx, IntConstant.v (0)), targetUnit) & if (vx == 0) goto target; \\ \hline
0x39 & if-nez vx,target & Checks vx and jumps if vx is nonzero2.  & 3902 1200 - if-nez v2, 0014 // +0012 Jumps to current position+18 words (hex 12) if v2 is nonzero. 0014 is the label of the target instruction. & newIfStmt (newNeExpr (vx, IntConstant.v (0)), targetUnit) & if (vx != 0) goto target; \\ \hline
0x3A & if-ltz vx,target & Checks vx and jumps if vx<02.  & 3A00 1600 - if-ltz v0, 002d // +0016 Jumps to the current position+16H words if v0<0. 002d is the label of the target instruction.  & newIfStmt (newLtExpr (vx, IntConstant.v (0)), targetUnit) & if (vx $<$ 0) goto target; \\ \hline
0x3B & if-gez vx,target & Checks vx and jumps if vx>=02. & 3B00 1600 - if-gez v0, 002d // +0016 Jumps to the current position+16H words if v0 >=0. 002d is the label of the target instruction. & newIfStmt (newGeExpr (vx, IntConstant.v (0)), targetUnit) & if (vx $>$= 0) goto target; \\ \hline
0x3C & if-gtz vx,target & Checks vx and jumps if vx>02.  & 3C00 1D00 - if-gtz v0, 004a // +001d Jumps to the current position+1DH words if v0>0. 004A is the label of the target instruction.  & newIfStmt (newGtExpr (vx, IntConstant.v (0)), targetUnit) & if (vx $>$ 0) goto target; \\ \hline
0x3D & if-lez vx,target & Checks vx and jumps if vx<=02.  & 3D00 1D00 - if-lez v0, 004a // +001d Jumps to the current position+1DH words if v0<=0. 004A is the label of the target instruction. & newIfStmt (newLeExpr (vx, IntConstant.v (0)), targetUnit) & if (vx $<$= 0) goto target; \\ \hline
0x3E & unused\_3E &  & & - & - \\ \hline
0x3F & unused\_3F &  & & - & - \\ \hline
0x40 & unused\_40 &  & & - & - \\ \hline
0x41 & unused\_41 &  & & - & - \\ \hline
0x42 & unused\_42 &  & & - & - \\ \hline
0x43 & unused\_43 &  & & - & - \\ \hline
0x44 & aget vx,vy,vz & Gets an integer value of an object reference array into vx. The array is referenced by vy and is indexed by vz.  & 4407 0306 - aget v7, v3, v6 Gets an integer array element. The array is referenced by v3 and the element is indexed by v6. The element will be put into v7. & newAssignStmt (vx, newArrayRef (vy, vz))& vx = vy[vz] \\ \hline
0x45 & aget-wide vx,vy,vz & Gets a long/double value of long/double array into vx,vx+1. The array is referenced by vy and is indexed by vz.  & 4505 0104 - aget-wide v5, v1, v4 Gets a long/double array element. The array is referenced by v1 and the element is indexed by v4. The element will be put into v5,v6. & newAssignStmt (vx, newArrayRef (vy, vz)) & vx = vy[vz] \\ \hline
0x46 & aget-object vx,vy,vz & Gets an object reference value of an object reference array into vx. The array is referenced by vy and is indexed by vz. & 4602 0200 - aget-object v2, v2, v0 Gets an object reference array element. The array is referenced by v2 and the element is indexed by v0. The element will be put into v2. & newAssignStmt (vx, newArrayRef (vy, vz)) & vx = vy[vz] \\ \hline
0x47 & aget-boolean vx,vy,vz & Gets a boolean value of a boolean array into vx. The array is referenced by vy and is indexed by vz.  & 4700 0001 - aget-boolean v0, v0, v1  Gets a boolean array element. The array is referenced by v0 and the element is indexed by v1. The element will be put into v0.  & newAssignStmt (vx, newArrayRef (vy, vz)) & vx = vy[vz] \\ \hline
0x48 & aget-byte vx,vy,vz & Gets a byte value of a byte array into vx. The array is referenced by vy and is indexed by vz.  & 4800 0001 - aget-byte v0, v0, v1 Gets a byte array element. The array is referenced by v0 and the element is indexed by v1. The element will be put into v0. & newAssignStmt (vx, newArrayRef (vy, vz)) & vx = vy[vz] \\ \hline
0x49 & aget-char vx, vy,vz & Gets a char value  of a character array into vx. The element is indexed by vz, the array object is referenced by vy  & 4905 0003 - aget-char v5, v0, v3 Gets a character array element. The array is referenced by v0 and the element is indexed by v3. The element will be put into v5. & newAssignStmt (vx, newArrayRef (vy, vz)) & vx = vy[vz] \\ \hline
0x4A & aget-short vx,vy,vz & Gets a short value  of a short array into vx. The element is indexed by vz, the array object is referenced by vy.  & 4A00 0001 - aget-short v0, v0, v1 Gets a short array element. The array is referenced by v0 and the element is indexed by v1. The element will be put into v0. & newAssignStmt (vx, newArrayRef (vy, vz)) & vx = vy[vz] \\ \hline
0x4B & aput vx,vy,vz & Puts the integer value in vx into an element of an integer array. The element is indexed by vz, the array object is referenced by vy.  & 4B00 0305 - aput v0, v3, v5 Puts the integer value in v2 into an integer array referenced by v0. The target array element is indexed by v1. & newAssignStmt (newArrayRef (vy, vz), vx) & vy[vz] = vx \\ \hline
0x4C & aput-wide vx,vy,vz & Puts the double/long value in vx,vx+1 into a double/long array. The array is referenced by vy, the element is indexed by vz. & 4C05 0104 - aput-wide v5, v1, v4 Puts the double/long value in v5,v6 into a double/long array referenced by v1. The target array element is indexed by v4. & newAssignStmt (newArrayRef (vy, vz), vx) & vy[vz] = vx \\ \hline
0x4D & aput-object vx,vy,vz & Puts the object reference value in vx into an element of an object reference array. The element is indexed by vz, the array object is referenced by vy.  & 4D02 0100 - aput-object v2, v1, v0 Puts the object reference value in v2 into an object reference array referenced by v0. The target array element is indexed by v1. & newAssignStmt (newArrayRef (vy, vz), vx) & vy[vz] = vx \\ \hline
0x4E & aput-boolean vx,vy,vz & Puts the boolean value in vx into an element of a boolean array. The element is indexed by vz, the array object is referenced by vy.  & 4E01 0002 - aput-boolean v1, v0, v2 Puts the boolean value in v1 into an object reference array referenced by v0. The target array element is indexed by v2. & newAssignStmt (newArrayRef (vy, vz), vx) & vy[vz] = vx \\ \hline
0x4F & aput-byte vx,vy,vz & Puts the byte value in vx into an element of a byte array. The element is indexed by vz, the array object is referenced by vy.  & 4F02 0001 - aput-byte v2, v0, v1 Puts the boolean value in v2 into a byte array referenced by v0. The target array element is indexed by v1. & newAssignStmt (newArrayRef (vy, vz), vx) & vy[vz] = vx \\ \hline
0x50 & aput-char vx,vy,vz & Puts the char value in vx into an element of a character array. The element is indexed by vz, the array object is referenced by vy.  & 5003 0001 - aput-char v3, v0, v1 Puts the character value in v3 into a character array referenced by v0. The target array element is indexed by v1.  & newAssignStmt (newArrayRef (vy, vz), vx) & vy[vz] = vx \\ \hline
0x51 & aput-short vx,vy,vz & Puts the short value in vx into an element of a short array. The element is indexed by vz, the array object is referenced by vy.  & 5102 0001 - aput-short v2, v0, v1 Puts the short value in v2 into a character array referenced by v0. The target array element is indexed by v1. & newAssignStmt (newArrayRef (vy, vz), vx) & vy[vz] = vx \\ \hline
0x52 & iget vx, vy, field\_id & Reads an instance field into vx. The instance is referenced by vy. & 5210 0300 - iget v0, v1, Test2.i6:I // field@0003 Reads field@0003 into v0 (entry \#3 in the field id table). The instance is referenced by v1. & & vx = field\_id \\ \hline
0x53 & iget-wide vx,vy,field\_id & Reads an instance field into vx1. The instance is referenced by vy. & 5320 0400 - iget-wide v0, v2, Test2.l0:J // field@0004 Reads field@0004 into v0 and v1 registers (entry \#4 in the field id table). The instance is referenced by v2. & & vx = field\_id \\ \hline
0x54 & iget-object vx,vy,field\_id & Reads an object reference instance field into vx. The instance is referenced by vy. & iget-object v1, v2, LineReader.fis:Ljava/io/FileInputStream; // field@0002 Reads field@0002 into v1  (entry \#2 in the field id table). The instance is referenced by v2. & & vx = field\_id \\ \hline
0x55 & iget-boolean vx,vy,field\_id & Reads a boolean instance field into vx. The instance is referenced by vy. & 55FC 0000 - iget-boolean v12, v15, Test2.b0:Z // field@0000 Reads the boolean field@0000 into v12 register (entry \#0 in the field id table). The instance is referenced by v15. & & vx = field\_id \\ \hline
0x56 & iget-byte vx,vy,field\_id & Reads a byte instance field into vx. The instance is referenced by vy.  & 5632 0100 - iget-byte v2, v3, Test3.bi1:B // field@0001 Reads the char field@0001 into v2 register (entry \#1 in the field id table). The instance is referenced by v3. & & vx = field\_id \\ \hline
0x57 & iget-char vx,vy,field\_id & Reads a char instance field into vx. The instance is referenced by vy.  & 5720 0300 - iget-char v0, v2, Test3.ci1:C // field@0003 Reads the char field@0003 into v0 register (entry \#3 in the field id table). The instance is referenced by v2. & & vx = field\_id \\ \hline
0x58 & iget-short vx,vy,field\_id & Reads a short instance field into vx. The instance is referenced by vy.  & 5830 0800 - iget-short v0, v3, Test3.si1:S // field@0008 Reads the short field@0008 into v0 register (entry \#8 in the field id table). The instance is referenced by v3. & & vx = field\_id \\ \hline
0x59 & iput vx,vy, field\_id & Puts vx into an instance field. The instance is referenced by vy. & 5920 0200 - iput v0,v2, Test2.i6:I // field@0002 Stores v0 into field@0002 (entry \#2 in the field id table). The instance is referenced by v2. & & field\_id = vx \\ \hline
0x5A & iput-wide vx,vy, field\_id & Puts the wide value located in vx and vx+1 registers into an instance field. The instance is referenced by vy. & 5A20 0000 - iput-wide v0,v2, Test2.d0:D // field@0000  Stores the wide value in v0, v1 registers into field@0000 (entry \#0 in the field id table). The instance is referenced by v2. & & field\_id = vx \\ \hline
0x5B & iput-object vx,vy,field\_id & Puts the object reference in vx into an instance field. The instance is referenced by vy. & 5B20 0000 - iput-object v0, v2, LineReader.bis:Ljava/io/BufferedInputStream; // field@0000 Stores the object reference in v0 into field@0000 (entry \#0 in the field table). The instance is referenced by v2. & & field\_id = vx \\ \hline
0x5C & iput-boolean vx,vy, field\_id & Puts the boolean value located in vx into an instance field. The instance is referenced by vy. & 5C30 0000 - iput-boolean v0, v3, Test2.b0:Z // field@0000 Puts the boolean value in v0 into field@0000 (entry \#0 in the field id table). The instance is referenced by v3. & & field\_id = vx \\ \hline
0x5D & iput-byte vx,vy,field\_id & Puts the byte value located in vx into an instance field. The instance is referenced by vy.  & 5D20 0100 - iput-byte v0, v2, Test3.bi1:B // field@0001 Puts the boolean value in v0 into field@0001 (entry \#1 in the field id table). The instance is referenced by v2. & & field\_id = vx \\ \hline
0x5E & iput-char vx,vy,field\_id & Puts the char value located in vx into an instance field. The instance is referenced by vy.  & 5E20 0300 - iput-char v0, v2, Test3.ci1:C // field@0003 Puts the char value in v0 into field@0003 (entry \#3 in the field id table). The instance is referenced by v2. & & field\_id = vx \\ \hline
0x5F & iput-short vx,vy,field\_id & Puts the short value located in vx into an instance field. The instance is referenced by vy.  & 5F21 0800 - iput-short v1, v2, Test3.si1:S // field@0008 Puts the short value in v1 into field@0008 (entry \#8 in the field id table). The instance is referenced by v2. & & field\_id = vx \\ \hline
0x60 & sget vx,field\_id & Reads the integer field identified by the field\_id into vx.  & 6000 0700 - sget v0, Test3.is1:I // field@0007 Reads field@0007 (entry \#7 in the field id table) into v0. & & vx = field\_id \\ \hline
0x61 & sget-wide vx, field\_id & Reads the static field identified by the field\_id into vx and vx+1 registers. & 6100 0500 - sget-wide v0, Test2.l1:J // field@0005 Reads field@0005 (entry \#5 in the field id table) into v0 and v1 registers. & & vx = field\_id \\ \hline
0x62 & sget-object vx,field\_id & Reads the object reference field identified by the field\_id into vx.  & 6201 0C00 - sget-object v1, Test3.os1:Ljava/lang/Object; // field@000c Reads field@000c (entry \#CH in the field id table) into v1. & & vx = field\_id \\ \hline
0x63 & sget-boolean vx,field\_id & Reads the boolean static field identified by the field\_id into vx. & 6300 0C00 - sget-boolean v0, Test2.sb:Z // field@000c Reads boolean field@000c (entry \#12 in the field id table) into v0. & & vx = field\_id \\ \hline
0x64 & sget-byte vx,field\_id & Reads the byte static field identified by the field\_id into vx.  & 6400 0200 - sget-byte v0, Test3.bs1:B // field@0002 Reads byte field@0002 (entry \#2 in the field id table) into v0. & & vx = field\_id \\ \hline
0x65 & sget-char vx,field\_id & Reads the char static field identified by the field\_id into vx.  & 6500 0700 - sget-char v0, Test3.cs1:C // field@0007 Reads byte field@0007 (entry \#7 in the field id table) into v0. & & vx = field\_id \\ \hline
0x66 & sget-short vx,field\_id & Reads the short static field identified by the field\_id into vx.  & 6600 0B00 - sget-short v0, Test3.ss1:S // field@000b Reads short field@000b (entry \#BH in the field id table) into v0. & & vx = field\_id \\ \hline
0x67 & sput vx, field\_id & Puts vx into a static field. & 6700 0100 - sput v0, Test2.i5:I // field@0001 Stores v0 into field@0001 (entry \#1 in the field id table). & & field\_id = vx \\ \hline
0x68 & sput-wide vx, field\_id & Puts vx and vx+1 into a static field. & 6800 0500 - sput-wide v0, Test2.l1:J // field@0005 Puts the long value in v0 and v1 into the field@0005 static field (entry \#5 in the field id table). & & field\_id = vx \\ \hline
0x69 & sput-object vx,field\_id & Puts object reference in vx into a static field.  & 6900 0c00 - sput-object v0, Test3.os1:Ljava/lang/Object; // field@000c Puts the object reference value in v0 into the field@000c static field (entry \#CH in the field id table). & & field\_id = vx \\ \hline
0x6A & sput-boolean vx,field\_id & Puts boolean value in vx into a static field.  & 6A00 0300 - sput-boolean v0, Test3.bls1:Z // field@0003 Puts the byte value in v0 into the field@0003 static field (entry \#3 in the field id table). & & field\_id = vx \\ \hline
0x6B & sput-byte vx,field\_id & Puts byte value in vx into a static field.  & 6B00 0200 - sput-byte v0, Test3.bs1:B // field@0002 Puts the byte value in v0 into the field@0002 static field (entry \#2 in the field id table). & & field\_id = vx \\ \hline
0x6C & sput-char vx,field\_id & Puts char value in vx into a static field.  & 6C01 0700 - sput-char v1, Test3.cs1:C // field@0007 Puts the char value in v1 into the field@0007 static field (entry \#7 in the field id table). & & field\_id = vx \\ \hline
0x6D & sput-short vx,field\_id & Puts short value in vx into a static field.  & 6D00 0B00 - sput-short v0, Test3.ss1:S // field@000b Puts the short value in v0 into the field@000b static field (entry \#BH in the field id table). & & field\_id = vx \\ \hline
0x6E & invoke-virtual \{ parameters \}, methodtocall & Invokes a virtual method with parameters. & 6E53 0600 0421 - invoke-virtual \{ v4, v0, v1, v2, v3\}, Test2.method5:(IIII)V // method@0006 Invokes the 6th method in the method table with the following arguments: v4 is the ""this"" instance, v0, v1, v2, and v3 are the method parameters. The method has 5 arguments (4 MSB bits of the second byte)5. & & invoke-virtual \\ \hline
0x6F & invoke-super \{parameter\},methodtocall & Invokes the virtual method of the immediate parent class.  & 6F10 A601 0100 invoke-super \{v1\},java.io.FilterOutputStream.close:()V // method@01a6 Invokes method@01a6 with one parameter, v1. & & invoke-special \\ \hline
0x70 & invoke-direct \{ parameters \}, methodtocall & Invokes a method with parameters without the virtual method resolution. & 7010 0800 0100 - invoke-direct \{v1\}, java.lang.Object.<init>:()V // method@0008 Invokes the 8th method in the method table with just one parameter, v1 is the ""this"" instance5. & & invoke-special \\ \hline
0x71 & invoke-static \{parameters\}, methodtocall & Invokes a static method with parameters. & 7110 3400 0400 - invoke-static \{v4\}, java.lang.Integer.parseInt:( Ljava/lang/String;)I // method@0034 Invokes method@34 static method. The method is called with one parameter, v45. & & invoke-static \\ \hline
0x72 & invoke-interface \{parameters\},methodtocall & Invokes an interface method. & 7240 2102 3154 invoke-interface \{v1, v3, v4, v5\}, mwfw.IReceivingProtocolAdapter.receivePackage:( ILjava/lang/String;Ljava/io/InputStream;)Z // method@0221 Invokes method@221 interface method using parameters in v1,v3,v4 and v55. & & invoke-interface \\ \hline
0x73 & unused\_73 &  & & & - \\ \hline
0x74 & invoke-virtual/range \{vx..vy\},methodtocall & Invokes virtual method with a range of registers. The instruction specifies the first register and the number of registers to be passed to the method.  & 7403 0600 1300 - invoke-virtual \{v19..v21\}, Test2.method5:(IIII)V // method@0006 Invokes the 6th method in the method table with the following arguments: v19 is the ""this"" instance, v20 and v21 are the method parameters.   & & invoke-virtual \\ \hline
0x75 & invoke-super/range & Invokes  the virtual method of the immediate parent class. The instruction specifies the first register and the number of registers to be passed to the method.  & 7501 A601 0100 invoke-super \{v1\},java.io.FilterOutputStream.close:()V // method@01a6 Invokes method@01a6 with one parameter, v1.  & & invoke-special \\ \hline
0x76 & invoke-direct/range \{vx..vy\},methodtocall & Invokes direct method with a range of registers. The instruction specifies the first register and the number of registers to be passed to the method. & 7603 3A00 1300 - invoke-direct/range \{v19..21\},java.lang.Object.<init>:()V // method@003a Invokes method@3A with 1 parameters (second byte of the instruction=03). The parameter is stored in v19 (5th,6th bytes of the instruction). & & invoke-special \\ \hline
0x77 & invoke-static/range \{vx..vy\},methodtocall & Invokes static method with a range of registers. The instruction specifies the first register and the number of registers to be passed to the method.  & 7703 3A00 1300 - invoke-static/range \{v19..21\},java.lang.Integer.parseInt:( Ljava/lang/String;)I // method@0034 Invokes method@3A with 1 parameters (second byte of the instruction=03). The parameter is stored in v19 (5th,6th bytes of the instruction).  & & invoke-static \\ \hline
0x78 & invoke-interface-range & Invokes an interface method with a range of registers. The instruction specifies the first register and the number of registers to be passed to the method.  & 7840 2102 0100 invoke-interface \{v1..v4\}, mwfw.IReceivingProtocolAdapter.receivePackage:( ILjava/lang/String;Ljava/io/InputStream;)Z // method@0221 Invokes method@221 interface method using parameters in v1..v4.  & & invoke-interface \\ \hline
0x79 & unused\_79 &  & & - & - \\ \hline
0x7A & unused\_7A &  & & - & - \\ \hline
0x7B & neg-int vx,vy & Calculates vx=-vy. & 7B01 - neg-int v1,v0 Calculates -v0 and stores the result in v1. & newAssignStmt ( vx, newNegExpr (vy)) & vx = 0 -vy \\ \hline
0x7C & not-int vx,vy &  & & ? & vx = vy \^{} (-1) \\ \hline
0x7D & neg-long vx,vy & Calculates vx,vx+1=-(vy,vy+1)  & 7D02 - neg-long v2,v0 Calculates -(v0,v1) and stores the result into (v2,v3) & newAssignStmt ( vx, newNegExpr (vy)) & vx = 0 -vy \\ \hline
0x7E & not-long vx,vy &  & & ? & vx = vy \^{} (-1) \\ \hline
0x7F & neg-float vx,vy & Calculates vx=-vy & 7F01 - neg-float v1,v0 Calculates -v0 and stores the result into v1. & newAssignStmt ( vx, newNegExpr (vy)) & vx = 0 -vy \\ \hline
0x80 & neg-double vx,vy & Calculates vx,vx+1=-(vy,vy+1)  & 8002 - neg-double v2,v0 Calculates -(v0,v1) and stores the result into (v2,v3) & newAssignStmt ( vx, newNegExpr (vy)) & vx = 0 -vy \\ \hline
0x81 & int-to-long vx, vy & Converts the integer in vy into a long in vx,vx+1. & 8106 - int-to-long v6, v0 Converts an integer in v0 into a long in v6,v7. & newAssignStmt (vx, newCastExpr (vy, LongType)) & vx = (long) vy \\ \hline
0x82 & int-to-float vx, vy & Converts the integer in vx into a float in vx. & 8206 - int-to-float v6, v0 Converts the integer in v0 into a float in v6. & newAssignStmt (vx, newCastExpr (vy, FloatType)) & vx = (float) vy \\ \hline
0x83 & int-to-double vx, vy & Converts the integer in vy into the double in vx,vx+1. & 8306 - int-to-double v6, v0 Converts the integer in v0 into a double in v6,v7 & newAssignStmt (vx, newCastExpr (vy, DoubleType)) & vx = (double) vy \\ \hline
0x84 & long-to-int vx,vy & Converts the long value in vy,vy+1 into an integer in vx. & 8424 - long-to-int v4, v2 Converts the long value in v2,v3 into an integer value in v4. & newAssignStmt (vx, newCastExpr (vy, IntType)) & vx = (int) vy \\ \hline
0x85 & long-to-float vx, vy & Converts the long value in vy,vy+1 into a float in vx. & 8510 - long-to-float v0, v1 Convcerts the long value in v1,v2 into a float value in v0. & newAssignStmt (vx, newCastExpr (vy, FloatType)) & vx = (float) vy \\ \hline
0x86 & long-to-double vx, vy & Converts the long value in vy,vy+1 into a double value in vx,vx+1. & 8610 - long-to-double v0, v1 Converts the long value in v1,v2 into a double value in v0,v1. & newAssignStmt (vx, newCastExpr (vy, DoubleType)) & vx = (double) vy \\ \hline
0x87 & float-to-int vx, vy & Converts the float value in vy into an integer value in vx. & 8730 - float-to-int v0, v3 Converts the float value in v3 into an integer value in v0. & newAssignStmt (vx, newCastExpr (vy, IntType)) & vx = (int) vy \\ \hline
0x88 & float-to-long vx,vy & Converts the float value in vy into a long value in vx. & 8830 - float-to-long v0, v3 Converts the float value in v3 into a long value in v0,v1. & newAssignStmt (vx, newCastExpr (vy, LongType)) & vx = (long) vy \\ \hline
0x89 & float-to-double vx, vy & Converts the float value in vy into a double value in vx,vx+1. & 8930 - float-to-double v0, v3 Converts the float value in v3 into a double value in v0,v1. & newAssignStmt (vx, newCastExpr (vy, DoubleType)) & vx = (double) vy \\ \hline
0x8A & double-to-int vx, vy & Converts the double value in vy,vy+1 into an integer value in vx. & 8A40  - double-to-int v0, v4 Converts the double value in v4,v5 into an integer value in v0. & newAssignStmt (vx, newCastExpr (vy, IntType)) & vx = (int) vy \\ \hline
0x8B & double-to-long vx, vy & Converts the double value in vy,vy+1 into a long value in vx,vx+1. & 8B40 - double-to-long v0, v4 Converts the double value in v4,v5 into a long value in v0,v1. & newAssignStmt (vx, newCastExpr (vy, LongType)) & vx = (long) vy \\ \hline
0x8C & double-to-float vx, vy & Converts the double value in vy,vy+1 into a float value in vx. & 8C40 - double-to-float v0, v4 Converts the double value in v4,v5 into a float value in v0,v1. & newAssignStmt (vx, newCastExpr (vy, FloatType)) & vx = (float) vy \\ \hline
0x8D & int-to-byte vx,vy & Converts the int value in vy to a byte value and stores it in vx. & 8D00 - int-to-byte v0, v0 Converts the integer in v0 into a byte and puts the byte value into v0. & newAssignStmt (vx, newCastExpr (vy, ByteType)) & vx = (byte) vy \\ \hline
0x8E & int-to-char vx,vy & Converts the int value in vy to a char value and stores it in vx. & 8E33  - int-to-char v3, v3 Converts the integer in v3 into a char and puts the char value into v3. & newAssignStmt (vx, newCastExpr (vy, CharType)) & vx = (char) vy \\ \hline
0x8F & int-to-short vx,vy & Converts the int value in vy to a short value and stores it in vx.  & 8F00 - int-to-short v0, v0 Converts the integer in v0 into a short and puts the short value into v3. & newAssignStmt (vx, newCastExpr (vy, ShortType)) & vx = (short) vy \\ \hline
0x90 & add-int vx,vy,vz & Calculates vy+vz and puts the result into vx. & 9000 0203 - add-int v0, v2, v3 Adds v3 to v2 and puts the result into v04. & newAssignStmt (vx, newAddExpr (vy, vz)) & vx = vy + vz \\ \hline
0x91 & sub-int vx,vy,vz & Calculates vy-vz and puts the result into vx. & 9100 0203 - sub-int v0, v2, v3 Subtracts v3 from v2 and puts the result into v0. & newAssignStmt (vx, newSubExpr (vy, vz)) & vx = vy - vz \\ \hline
0x92 & mul-int vx, vy, vz & Multiplies vz with wy and puts the result int vx. & 9200 0203 - mul-int v0,v2,v3 Multiplies v2 with w3 and puts the result into v0 & newAssignStmt (vx, newMulExpr (vy, vz)) & vx = vy * vz \\ \hline
0x93 & div-int vx,vy,vz & Divides vy with vz and puts the result into vx. & 9303 0001 - div-int v3, v0, v1 Divides v0 with v1 and puts the result into v3. & newAssignStmt (vx, newDivExpr (vy, vz)) & vx = vy / vz \\ \hline
0x94 & rem-int vx,vy,vz & Calculates vy \% vz and puts the result into vx.  & 9400 0203 - rem-int v0, v2, v3 Calculates v3 \% v2 and puts the result into v0.  & newAssignStmt (vx, newRemExpr (vy, vz)) & vx = vy \% vz \\ \hline
0x95 & and-int vx, vy, vz & Calculates vy AND vz and puts the result into vx. & 9503 0001 - and-int v3, v0, v1 Calculates v0 AND v1 and puts the result into v3. & newAssignStmt (vx, newAndExpr (vy, vz)) & vx = vy \& vz \\ \hline
0x96 & or-int vx, vy, vz & Calculates vy OR vz and puts the result into vx. & 9603 0001 - or-int v3, v0, v1 Calculates v0 OR v1 and puts the result into v3. & newAssignStmt (vx, newOrExpr (vy, vz)) & vx = vy \textpipe{} vz \\ \hline
0x97 & xor-int vx, vy, vz & Calculates vy XOR vz and puts the result into vx. & 9703 0001 - xor-int v3, v0, v1 Calculates v0 XOR v1 and puts the result into v3. & newAssignStmt (vx, newXorExpr (vy, vz)) & vx = vy \^{} vz \\ \hline
0x98 & shl-int vx, vy, vz & Shift vy left by the positions specified by vz and store the result into vx. & 9802 0001 - shl-int v2, v0, v1 Shift v0 left by the positions specified by v1 and store the result in v2. & newAssignStmt (vx, newShlExpr (vy, vz)) & vx = vy $<<$ vz \\ \hline
0x99 & shr-int vx, vy, vz & Shift vy right by the positions specified by vz and store the result into vx. & 9902 0001 - shr-int v2, v0, v1 Shift v0 right by the positions specified by v1 and store the result in v2. & newAssignStmt (vx, newShrExpr (vy, vz)) & vx = vy $>>$ vz \\ \hline
0x9A & ushr-int vx, vy, vz & Unsigned shift right (>>>) vy by the positions specified by vz and store the result into vx. & 9A02 0001 - ushr-int v2, v0, v1 Unsigned shift v0 right by the positions specified by v1 and store the result in v2. & newAssignStmt (vx, newUshrExpr (vy, vz)) & vx = vy $>>$ vz \\ \hline
0x9B & add-long vx, vy, vz & Adds vy to vz and puts the result into vx1. & 9B00 0305 - add-long v0, v3, v5 The long value in v3,v4 is added to the value in v5,v6 and the result is stored in v0,v1. & newAssignStmt (vx, newAddExpr (vy, vz)) & vx = vy + vz \\ \hline
0x9C & sub-long vx,vy,vz & Calculates vy-vz and puts the result into vx1. & 9C00 0305 - sub-long v0, v3, v5 Subtracts the long value in v5,v6 from the long value in v3,v4 and puts the result into v0,v1. & newAssignStmt (vx, newSubExpr (vy, vz)) & vx = vy - vz \\ \hline
0x9D & mul-long vx,vy,vz & Calculates vy*vz and puts the result into vx1.  & 9D00 0305 - mul-long v0, v3, v5 Multiplies the long value in v5,v6 with the long value in v3,v4 and puts the result into v0,v1.  & newAssignStmt (vx, newMulExpr (vy, vz)) & vx = vy * vz \\ \hline
0x9E & div-long vx, vy, vz & Calculates vy/vz and puts the result into vx1. & 9E06 0002 - div-long v6, v0, v2 Divides the long value in v0,v1 with the long value in v2,v3 and pust the result into v6,v7. & newAssignStmt (vx, newDivExpr (vy, vz)) & vx = vy / vz \\ \hline
0x9F & rem-long vx,vy,vz & Calculates vy \% vz and puts the result into vx1.  & 9F06 0002 - rem-long v6, v0, v2 Calculates v0,v1 \%  v2,v3 and puts the result into v6,v7.  & newAssignStmt (vx, newRemExpr (vy, vz)) & vx = vy \% vz \\ \hline
0xA0 & and-long vx, vy, vz & Calculates the vy AND vz and puts the result into vx1. & A006 0002 - and-long v6, v0, v2 Calculates v0,v1 AND v2,v3 and puts the result into v6,v7. & newAssignStmt (vx, newAndExpr (vy, vz)) & vx = vy \& vz \\ \hline
0xA1 & or-long vx, vy, vz & Calculates the vy OR vz and puts the result into vx1. & A106 0002 - or-long v6, v0, v2 Calculates v0,v1 OR v2,v3 and puts the result into v6,v7. & newAssignStmt (vx, newOrExpr (vy, vz)) & vx = vy \textpipe{} vz \\ \hline
0xA2 & xor-long vx, vy, vz & Calculates the vy XOR vz and puts the result into vx1. & A206 0002 - xor-long v6, v0, v2 Calculates v0,v1 XOR v2,v3 and puts the result into v6,v7. & newAssignStmt (vx, newXorExpr (vy, vz)) & vx = vy \^{} vz \\ \hline
0xA3 & shl-long vx, vy, vz & Shifts left vy by vz positions and stores the result in vx1. & A302 0004 - shl-long v2, v0, v4 Shift v0,v1 by postions specified by v4 and puts the result into v2,v3. & newAssignStmt (vx, newShlExpr (vy, vz)) & vx = vy $<<$ vz \\ \hline
0xA4 & shr-long vx,vy,vz & Shifts right vy by vz positions and stores the result in vx1. & A402 0004 - shr-long v2, v0, v4 Shift v0,v1 by postions specified by v4 and puts the result into v2,v3. & newAssignStmt (vx, newShrExpr (vy, vz)) & vx = vy $>>$ vz \\ \hline
0xA5 & ushr-long vx, vy, vz & Unsigned shifts right vy by vz positions and stores the result in vx1. & A502 0004 - ushr-long v2, v0, v4 Unsigned shift v0,v1 by postions specified by v4 and puts the result into v2,v3. & newAssignStmt (vx, newUshrExpr (vy, vz)) & vx = vy $>>$ vz \\ \hline
0xA6 & add-float vx,vy,vz & Adds vy to vz and puts the result into vx. & A600 0203 - add-float v0, v2, v3 Adds the floating point numbers in v2 and v3 and puts the result into v0. & newAssignStmt (vx, newAddExpr (vy, vz)) & vx = vy + vz \\ \hline
0xA7 & sub-float vx,vy,vz & Calculates vy-vz and puts the result into vx.  & A700 0203 - sub-float v0, v2, v3 Calculates v2-v3 and puts the result into v0.  & newAssignStmt (vx, newSubExpr (vy, vz)) & vx = vy - vz \\ \hline
0xA8 & mul-float vx, vy, vz & Multiplies vy with vz and puts the result into vx. & A803 0001 - mul-float v3, v0, v1 Multiplies v0 with v1 and puts the result into v3. & newAssignStmt (vx, newMulExpr (vy, vz)) & vx = vy * vz \\ \hline
0xA9 & div-float vx, vy, vz & Calculates vy/vz and puts the result into vx. & A903 0001 - div-float v3, v0, v1 Divides v0 with v1 and puts the result into v3. & newAssignStmt (vx, newDivExpr (vy, vz)) & vx = vy / vz \\ \hline
0xAA & rem-float vx,vy,vz & Calculates vy \% vz and puts the result into vx.  & AA03 0001 - rem-float v3, v0, v1 Calculates v0 \%  v1 and puts the result into v3.  & newAssignStmt (vx, newRemExpr (vy, vz)) & vx = vy \% vz \\ \hline
0xAB & add-double vx,vy,vz & Adds vy to vz and puts the result into vx1.  & AB00 0305 - add-double v0, v3, v5 Adds the double value in v5,v6 registers to the double value in v3,v4 registers and places the result  in v0,v1 registers. & newAssignStmt (vx, newAddExpr (vy, vz)) & vx = vy + vz \\ \hline
0xAC & sub-double vx,vy,vz & Calculates vy-vz and puts the result into vx1. & AC00 0305 - sub-double v0, v3, v5 Subtracts the value in v5,v6 from the value in v3,v4 and puts the result into v0,v1. & newAssignStmt (vx, newSubExpr (vy, vz)) & vx = vy - vz \\ \hline
0xAD & mul-double vx, vy, vz & Multiplies vy with vz and puts the result into vx1. & AD06 0002 - mul-double v6, v0, v2 Multiplies the double value in v0,v1 with the double value in v2,v3 and puts the result into v6,v7. & newAssignStmt (vx, newMulExpr (vy, vz)) & vx = vy * vz \\ \hline
0xAE & div-double vx, vy, vz & Calculates vy/vz and puts the result into vx1. & AE06 0002 - div-double v6, v0, v2 Divides the double value in v0,v1 with the double value in v2,v3 and puts the result into v6,v7. & newAssignStmt (vx, newDivExpr (vy, vz))& vx = vy / vz \\ \hline
0xAF & rem-double vx,vy,vz & Calculates vy \% vz and puts the result into vx1.  & AF06 0002 - rem-double v6, v0, v2 Calculates v0,v1 \% v2,v3 and puts the result into v6,v7.  & newAssignStmt (vx, newRemExpr (vy, vz)) & vx = vy \% vz \\ \hline
0xB0 & add-int/2addr vx,vy & Adds vy to vx. & B010 - add-int/2addr v0,v1 Adds v1 to v0. & newAssignStmt (vx, newAddExpr (vx, vy)) & vx = vx + vy \\ \hline
0xB1 & sub-int/2addr vx,vy & Calculates vx-vy and puts the result into vx. & B140 - sub-int/2addr v0, v4 Subtracts v4 from v0 and puts the result into v0. & newAssignStmt (vx, newSubExpr (vx, vy)) & vx = vx - vy \\ \hline
0xB2 & mul-int/2addr vx,vy & Multiplies vx with vy. & B210 - mul-int/2addr v0, v1 Multiples v0 with v1 and puts the result into v0. & newAssignStmt (vx, newMulExpr (vx, vy)) & vx = vx * vy \\ \hline
0xB3 & div-int/2addr vx,vy & Divides vx with vy and puts the result into vx. & B310 - div-int/2addr v0, v1 Divides v0 with v1 and puts the result into v0. & newAssignStmt (vx, newDivExpr (vx, vy)) & vx = vx / vy \\ \hline
0xB4 & rem-int/2addr vx,vy & Calculates vx \% vy and puts the result into vx  & B410 - rem-int/2addr v0, v1  Calculates v0 \% v1 and puts the result into v0.  & newAssignStmt (vx, newRemExpr (vx, vy)) & vx = vx \% vy \\ \hline
0xB5 & and-int/2addr vx, vy & Calculates vx AND vy and puts the result into vx. & B510 - and-int/2addr v0, v1 Calculates v0 AND v1 and puts the result into v0. & newAssignStmt (vx, newAndExpr (vx, vy)) & vx = vx \& vy \\ \hline
0xB6 & or-int/2addr vx, vy & Calculates vx OR vy and puts the result into vx. & B610 - or-int/2addr v0, v1 Calculates v0 OR v1 and puts the result into v0. & newAssignStmt (vx, newOrExpr (vx, vy)) & vx = vx \textpipe{} vy \\ \hline
0xB7 & xor-int/2addr vx, vy & Calculates vx XOR vy and puts the result into vx. & B710  - xor-int/2addr v0, v1 Calculates v0 XOR v1 and puts the result into v0. & newAssignStmt (vx, newXorExpr (vx, vy)) & vx = vx \^{} vy \\ \hline
0xB8 & shl-int/2addr vx, vy & Shifts vx left by vy positions. & B810 - shl-int/2addr v0, v1 Shift v0 left by v1 positions. & newAssignStmt (vx, newShlExpr (vx, vy)) & vx = vx $<<$ vy \\ \hline
0xB9 & shr-int/2addr vx, vy & Shifts vx right by vy positions. & B910 - shr-int/2addr v0, v1 Shift v0 right by v1 positions. & newAssignStmt (vx, newShrExpr (vx, vy)) & vx = vx $>>$ vy \\ \hline
0xBA & ushr-int/2addr vx, vy & Unsigned shift right (>>>) vx by the positions specified by vy. & BA10 - ushr-int/2addr v0, v1 Unsigned shift v0 by the positions specified by v1. & newAssignStmt (vx, newUshrExpr (vx, vy)) & vx = vx $>>$ vy \\ \hline
0xBB & add-long/2addr vx,vy & Adds vy to vx1. & BB20 - add-long/2addr v0, v2 Adds the long value in v2,v3 registers to the long value in v0,v1 registers. & newAssignStmt (vx, newAddExpr (vx, vy)) & vx = vx + vy \\ \hline
0xBC & sub-long/2addr vx,vy & Calculates vx-vy and puts the result into vx1. & BC70 - sub-long/2addr v0, v7 Subtracts the long value in v7,v8 from the long value in v0,v1 and puts the result into v0,v1. & newAssignStmt (vx, newSubExpr (vx, vy)) & vx = vx - vy \\ \hline
0xBD & mul-long/2addr vx,vy & Calculates vx*vy and puts the result into vx1.  & BD70 - mul-long/2addr v0, v7 Multiplies the long value in v7,v8 with the long value in v0,v1 and puts the result into v0,v1.  & newAssignStmt (vx, newMulExpr (vx, vy)) & vx = vx * vy \\ \hline
0xBE & div-long/2addr vx, vy & Calculates vx/vy and puts the result into vx1. & BE20 - div-long/2addr v0, v2 Divides the long value in v0,v1 with the long value in v2,v3 and puts the result into v0,v1 & newAssignStmt (vx, newDivExpr (vx, vy)) & vx = vx / vy \\ \hline
0xBF & rem-long/2addr vx,vy & Calculates vx \% vy and puts the result into vx1.  & BF20 - rem-long/2addr v0, v2 Calculates v0,v1 \% v2,v3 and puts the result into v0,v1  & newAssignStmt (vx, newRemExpr (vx, vy)) & vx = vx \% vy \\ \hline
0xC0 & and-long/2addr vx, vy & Calculates vx AND vy and puts the result into vx1. & C020 - and-long/2addr v0, v2 Calculates v0,v1 OR v2,v3 and puts the result into v0,v1. & newAssignStmt (vx, newAndExpr (vx, vy)) & vx = vx \& vy \\ \hline
0xC1 & or-long/2addr vx, vy & Calculates vx OR vy and puts the result into vx1. & C120  - or-long/2addr v0, v2 Calculates v0,v1 OR v2,v3 and puts the result into v0,v1. & newAssignStmt (vx, newOrExpr (vx, vy)) & vx = vx \textpipe{} vy \\ \hline
0xC2 & xor-long/2addr vx, vy & Calculates vx XOR vy and puts the result into vx1. & C220 - xor-long/2addr v0, v2 Calculates v0,v1 XOR v2,v3 and puts the result into v0,v1. & newAssignStmt (vx, newXorExpr (vx, vy)) & vx = vx \^{} vy \\ \hline
0xC3 & shl-long/2addr vx, vy & Shifts left the value in vx,vx+1 by the positions specified by vy and stores the result in vx,vx+1. & C320 - shl-long/2addr v0, v2 Shifts left v0,v1 by the positions specified by v2. & newAssignStmt (vx, newShlExpr (vx, vy)) & vx = vx $<<$ vy \\ \hline
0xC4 & shr-long/2addr vx, vy & Shifts right the value in vx,vx+1 by the positions specified by vy and stores the result in vx,vx+1. & C420 - shr-long/2addr v0, v2 Shifts right v0,v1 by the positions specified by v2. & newAssignStmt (vx, newShrExpr (vx, vy)) & vx = vx $>>$ vy \\ \hline
0xC5 & ushr-long/2addr vx, vy & Unsigned shifts right the value in vx,vx+1 by the positions specified by vy and stores the result in vx,vx+1. & C520 - ushr-long/2addr v0, v2 Unsigned shifts right v0,v1 by the positions specified by v2. & newAssignStmt (vx, newUshrExpr (vx, vy)) & vx = vx $>>$ vy \\ \hline
0xC6 & add-float/2addr vx,vy & Adds vy to vx.  & C640 - add-float/2addr v0,v4 Adds v4 to v0. & newAssignStmt (vx, newAddExpr (vx, vy)) & vx = vx + vy \\ \hline
0xC7 & sub-float/2addr vx,vy & Calculates vx-vy and stores the result in vx.  & C740 - sub-float/2addr v0,v4 Adds v4 to v0.  & newAssignStmt (vx, newSubExpr (vx, vy)) & vx = vx - vy \\ \hline
0xC8 & mul-float/2addr vx, vy & Multiplies vx with vy. & C810 - mul-float/2addr v0, v1 Multiplies v0 with v1. & newAssignStmt (vx, newMulExpr (vx, vy)) & vx = vx * vy \\ \hline
0xC9 & div-float/2addr vx, vy & Calculates vx/vy and puts the result into vx. & C910 - div-float/2addr v0, v1 Divides v0 with v1 and puts the result into v0. & newAssignStmt (vx, newDivExpr (vx, vy)) & vx = vx / vy \\ \hline
0xCA & rem-float/2addr vx,vy & Calculates vx/vy and puts the result into vx.  & CA10 - rem-float/2addr v0, v1  Calculates v0 \% v1 and puts the result into v0.  & newAssignStmt (vx, newRemExpr (vx, vy)) & vx = vx \% vy \\ \hline
0xCB & add-double/2addr vx, vy & Adds vy to vx1. & CB70 - add-double/2addr v0, v7 Adds v7 to v0. & newAssignStmt (vx, newAddExpr (vx, vy)) & vx = vx + vy \\ \hline
0xCC & sub-double/2addr vx, vy & Calculates vx-vy and puts the result into vx1. & CC70 - sub-double/2addr v0, v7 Subtracts the value in v7,v8 from the value in v0,v1 and puts the result into v0,v1. & newAssignStmt (vx, newSubExpr (vx, vy)) & vx = vx - vy \\ \hline
0xCD & mul-double/2addr vx, vy & Multiplies vx with vy1. & CD20 - mul-double/2addr v0, v2 Multiplies the double value in v0,v1 with the double value in v2,v3 and puts the result into v0,v1. & newAssignStmt (vx, newMulExpr (vx, vy))& vx = vx * vy \\ \hline
0xCE & div-double/2addr vx, vy & Calculates vx/vy and puts the result into vx1. & CE20 - div-double/2addr v0, v2 Divides the double value in v0,v1 with the double value in v2,v3 and puts the value into v0,v1. & newAssignStmt (vx, newDivExpr (vx, vy))& vx = vx / vy \\ \hline
0xCF & rem-double/2addr vx,vy & Calculates vx \% vy and puts the result into vx1.  & CF20 - rem-double/2addr v0, v2  Calculates  v0,v1 \%  v2,v3 and puts the value into v0,v1.  & newAssignStmt (vx, newRemExpr (vx, vy)) & vx = vx \% vy \\ \hline
0xD0 & add-int/lit16 vx,vy,lit16 & Adds vy to lit16 and stores the result into vx. & D001 D204 - add-int/lit16 v1, v0, \#int 1234 // \#04d2 Adds v0 to literal 1234 and stores the result into v1. & newAssignStmt (vx, newAddExpr (vy, lit16)) & vx = vy + lit16 \\ \hline
0xD1 & sub-int/lit16 vx,vy,lit16 & Calculates vy - lit16 and stores the result into vx. & D101 D204 - sub-int/lit16 v1, v0, \#int 1234 // \#04d2 Calculates v0 - literal 1234 and stores the result into v1. & newAssignStmt (vx, newSubExpr (vy, lit16)) & vx = vy - lit16 \\ \hline
0xD2 & mul-int/lit16 vx,vy,lit16 & Calculates vy * lit16 and stores the result into vx.  & D201 D204 - mul-int/lit16 v1, v0, \#int 1234 // \#04d2 Calculates v0 * literal 1234 and stores the result into v1.  & newAssignStmt (vx, newMulExpr (vy, lit16)) & vx = vy * lit16 \\ \hline
0xD3 & div-int/lit16 vx,vy,lit16 & Calculates vy / lit16 and stores the result into vx.  & D301 D204 - div-int/lit16 v1, v0, \#int 1234 // \#04d2 Calculates v0 / literal 1234 and stores the result into v1.  & newAssignStmt (vx, newDivExpr (vy, lit16)) & vx = vy / lit16 \\ \hline
0xD4 & rem-int/lit16 vx,vy,lit16 & Calculates vy \% lit16 and stores the result into vx.  & D401 D204 - rem-int/lit16 v1, v0, \#int 1234 // \#04d2 Calculates v0 \% literal 1234 and stores the result into v1.  & newAssignStmt (vx, newRemExpr (vy, lit16)) & vx = vy \% lit16 \\ \hline
0xD5 & and-int/lit16 vx,vy,lit16 & Calculates vy AND lit16 and stores the result into vx.  & D501 D204 - and-int/lit16 v1, v0, \#int 1234 // \#04d2 Calculates v0 AND literal 1234 and stores the result into v1.  & newAssignStmt (vx, newAndExpr (vy, lit16)) & vx = vy \& lit16 \\ \hline
0xD6 & or-int/lit16 vx,vy,lit16 & Calculates vy OR lit16 and stores the result into vx.  & D601 D204 - or-int/lit16 v1, v0, \#int 1234 // \#04d2 Calculates v0 OR literal 1234 and stores the result into v1.  & newAssignStmt (vx, newOrExpr (vy, lit16)) & vx = vy \textpipe{} lit16 \\ \hline
0xD7 & xor-int/lit16 vx,vy,lit16 & Calculates vy XOR lit16 and stores the result into vx.  & D701 D204 - xor-int/lit16 v1, v0, \#int 1234 // \#04d2 Calculates v0 XOR literal 1234 and stores the result into v1.  & newAssignStmt (vx, newXorExpr (vy, lit16)) & vx = vy \^{} lit16 \\ \hline
0xD8 & add-int/lit8 vx,vy,lit8 & Adds vy to lit8 and stores the result into vx. & D800 0201 - add-int/lit8 v0,v2, \#int1 Adds literal 1 to v2 and stores the result into v0. & newAssignStmt (vx, newAddExpr (vy, lit8)) & vx = vy + lit8 \\ \hline
0xD9 & sub-int/lit8 vx,vy,lit8 & Calculates vy-lit8 and stores the result into vx.  & D900 0201 - sub-int/lit8 v0,v2, \#int1 Calculates v2-1 and stores the result into v0. & newAssignStmt (vx, newSubExpr (vy, lit8))  & vx = vy - lit8 \\ \hline
0xDA & mul-int/lit-8 vx,vy,lit8 & Multiplies vy with lit8 8-bit literal constant and puts the result into vx. & DA00 0002 - mul-int/lit8 v0,v0, \#int2 Multiplies v0 with literal 2 and puts the result into v0. & newAssignStmt (vx, newMulExpr (vy, lit8)) & vx = vy * lit8 \\ \hline
0xDB & div-int/lit8 vx,vy,lit8 & Calculates vy/lit8 and stores the result into vx.  & DB00 0203 - mul-int/lit8 v0,v2, \#int3 Calculates v2/3 and stores the result into v0.  & newAssignStmt (vx, newDivExpr (vy, lit8)) & vx = vy / lit8 \\ \hline
0xDC & rem-int/lit8 vx,vy,lit8 & Calculates vy \% lit8 and stores the result into vx.  & DC00 0203 - rem-int/lit8 v0,v2, \#int3 Calculates v2 \% 3 and stores the result into v0.  & newAssignStmt (vx, newRemExpr (vy, lit8)) & vx = vy \% lit8 \\ \hline
0xDD & and-int/lit8 vx,vy,lit8 & Calculates vy AND lit8 and stores the result into vx.  & DD00 0203 - and-int/lit8 v0,v2, \#int3 Calculates v2 AND 3 and stores the result into v0.  & newAssignStmt (vx, newAndExpr (vy, lit8)) & vx = vy \& lit8 \\ \hline
0xDE & or-int/lit8 vx, vy, lit8 & Calculates vy OR lit8 and puts the result into vx. & DE00 0203 - or-int/lit8 v0, v2, \#int 3 Calculates v2 OR literal 3 and puts the result into v0. & newAssignStmt (vx, newOrExpr (vy, lit8)) & vx = vy \textpipe{} lit8 \\ \hline
0xDF & xor-int/lit8 vx, vy, lit8 & Calculates vy XOR lit8 and puts the result into vx. & DF00 0203     |  0008: xor-int/lit8 v0, v2, \#int 3 Calculates v2 XOR literal 3 and puts the result into v0. & newAssignStmt (vx, newXorExpr (vy, lit8)) & vx = vy \^{} lit8 \\ \hline
0xE0 & shl-int/lit8 vx, vy, lit8 & Shift v0 left by the bit positions specified by the literal constant and put the result into vx. & E001 0001 - shl-int/lit8 v1, v0, \#int 1 Shift v0 left by 1 position and put the result into v1. & newAssignStmt (vx, newShlExpr (vy, lit8)) & vx = vy + lit8 \\ \hline
0xE1 & shr-int/lit8 vx, vy, lit8 & Shift v0 right by the bit positions specified by the literal constant and put the result into vx. & E101 0001 - shr-int/lit8 v1, v0, \#int 1 Shift v0 right by 1 position and put the result into v1. & newAssignStmt (vx, newShrExpr (vy, lit8)) & vx = vy + lit8 \\ \hline
0xE2 & ushr-int/lit8 vx, vy, lit8 & Unsigned right shift of v0 (>>>) by the bit positions specified by the literal constant and put the result into vx. & E201 0001 - ushr-int/lit8 v1, v0, \#int 1 Unsigned shift v0 right by 1 position and put the result into v1. & newAssignStmt (vx, newUshrExpr (vy, lit8)) & vx = vy + lit8 \\ \hline
0xE3 & unused\_E3 &  & & - & - \\ \hline
0xE4 & unused\_E4 &  & & - & - \\ \hline
0xE5 & unused\_E5 &  & & - & - \\ \hline
0xE6 & unused\_E6 &  & & - & - \\ \hline
0xE7 & unused\_E7 &  & & - & - \\ \hline
0xE8 & unused\_E8 &  & & - & - \\ \hline
0xE9 & unused\_E9 &  & & - & - \\ \hline
0xEA & unused\_EA &  & & - & - \\ \hline
0xEB & unused\_EB &  & & - & - \\ \hline
0xEC & unused\_EC &  & & - & - \\ \hline
0xED & unused\_ED &  & & - & - \\ \hline
0xEE & execute-inline \{parameters\},inline ID & Executes the inline method identified by inline ID6. & EE20 0300 0100 - execute-inline \{v1, v0\}, inline \#0003 Executes inline method \#3 using v1 as ""this"" and passing one parameter in v0. & - odex only& odex \\ \hline
0xEF & unused\_EF &  & & -  & - \\ \hline
0xF0 & invoke-direct-empty & Stands as a placeholder for pruned empty methods like Object.<init>. This acts as nop during normal execution6. & F010 F608 0000 - invoke-direct-empty \{v0\}, Ljava/lang/Object;.<init>:()V // method@08f6 Replacement for the empty method java/lang/Object;<init>. & - odex only& odex \\ \hline
0xF1 & unused\_F1 &  & & - & - \\ \hline
0xF2 & iget-quick vx,vy,offset & Gets the value stored at offset in vy instance's data area to vx6.  & F221 1000 - iget-quick v1, v2, [obj+0010] Gets the value at offset 0CH of the instance pointed by v2 and stores the object reference in v1. & - odex only & odex \\ \hline
0xF3 & iget-wide-quick vx,vy,offset & Gets the object reference value stored at offset in vy instance's data area to vx,vx+16.  & F364 3001 - iget-wide-quick v4, v6, [obj+0130] Gets the value at offset 130H of the instance pointed by v6 and stores the object reference in v4,v5. & - odex only & odex \\ \hline
0xF4 & iget-object-quick vx,vy,offset & Gets the object reference value stored at offset in vy instance's data area to vx6. & F431 0C00 - iget-object-quick v1, v3, [obj+000c] Gets the object reference value at offset 0CH of the instance pointed by v3 and stores the object reference in v1. & - odex only & odex \\ \hline
0xF5 & iput-quick vx,vy,offset & Puts the value stored in vx to offset in vy instance's data area6.  & F521 1000  - iput-quick v1, v2, [obj+0010] Puts the object reference value in v1 to offset 10H of the instance pointed by v2. & - odex only & odex \\ \hline
0xF6 & iput-wide-quick vx,vy,offset & Puts the value stored in vx,vx+1 to offset in vy instance's data area6.  & F652 7001 - iput-wide-quick v2, v5, [obj+0170] Puts the value in v2,v3 to offset 170H of the instance pointed by v5. & - odex only& odex \\ \hline
0xF7 & iput-object-quick vx,vy,offset & Puts the object reference value stored in vx to offset in vy instance's data area to vx6.  & F701 4C00 - iput-object-quick v1, v0, [obj+004c] Puts the object reference value in v1 to offset 0CH of the instance pointed by v3. & - odex only & odex \\ \hline
0xF8 & invoke-virtual-quick \{parameters\},vtable offset & Invokes a virtual method using the vtable of the target object6. & F820 B800 CF00 - invoke-virtual-quick \{v15, v12\}, vtable \#00b8 Invokes a virtual method. The target object instance is pointed by v15 and vtable entry \#B8 points to the method to be called. v12 is a parameter to the method call. & - odex only & odex \\ \hline
0xF9 & invoke-virtual-quick/range \{parameter range\},vtable offset & Invokes a virtual method using the vtable of the target object6  & F906 1800 0000 - invoke-virtual-quick/range \{v0..v5\},vtable \#0018 Invokes a method using the vtable of the instance pointed by v0. v1..v5 registers are parameters to the method call. & - odex only & odex \\ \hline
0xFA & invoke-super-quick \{parameters\},vtable offset & Invokes a virtual method in the target object's immediate parent class using the vtable of that parent class6.  & FA40 8100 3254  - invoke-super-quick \{v2, v3, v4, v5\}, vtable \#0081 Invokes a method using the vtable of the immediate parent class of instance pointed by v2. v3, v4 and v5 registers are parameters to the method call. & - odex only& odex \\ \hline
0xFB & invoke-super-quick/range \{register range\},vtable offset & Invokes a virtual method in the target object's immediate parent class using the vtable of that parent class6. & F906 1B00 0000 - invoke-super-quick/range \{v0..v5\}, vtable \#001b Invokes a method using the vtable of the immediate parent class of instance pointed by v0. v1..v5 registers are parameters to the method call. & - odex only & odex \\ \hline
0xFC & unused\_FC &  & & - & - \\ \hline
0xFD & unused\_FD &  & & - & - \\ \hline
0xFE & unused\_FE &  & & - & - \\ \hline
0xFF & unused\_FF &  & & - & - \\ \hline